\documentclass[preprint]{aastex}

\usepackage{natbib}
\usepackage{lineno}
\usepackage[nomessages]{fp}
\newcommand{\nFGLh}{3646}

\newcommand{\nnofgl}{1353}
\newcommand{\sprobnew}{35.6}
\newcommand{\nsrc}{2863}
\newcommand{\probtot}{44.4}
\newcommand{\nbay}{2766}
\newcommand{\probbay}{34.8}
\newcommand{\nbayonly}{684}
\newcommand{\nlr}{2179}
\newcommand{\problr}{143.1}
\newcommand{\nlronly}{97}
\newcommand{\nnofglf}{204}
\newcommand{\nnofglb}{290}
\newcommand{\nnofglu}{822}
\newcommand{\nzz}{405}
\newcommand{\nnagn}{36}
\newcommand{\nhighlatf}{655}
\newcommand{\nhighlatbl}{1067}
\newcommand{\nhighlatu}{1077}
\newcommand{\nhighlatag}{64}
\newcommand{\nhighlataa}{587}
\newcommand{\nhighlatab}{11}
\newcommand{\nhighlatac}{5}
\newcommand{\nhighlatad}{52}
\newcommand{\nhighlatba}{297}
\newcommand{\nhighlatbb}{280}
\newcommand{\nhighlatbc}{326}
\newcommand{\nhighlatbd}{164}
\newcommand{\nhighlatca}{382}
\newcommand{\nhighlatcb}{142}
\newcommand{\nhighlatcc}{128}
\newcommand{\nhighlatcd}{425}

\newcommand{\nrdg}{38}
\newcommand{\nagn}{10}
\newcommand{\nnlsy}{9}
\newcommand{\ncss}{5}
\newcommand{\nssrq}{2}
\newcommand{\nbznf}{562}
\newcommand{\nbzf}{1909}
\newcommand{\ibznf}{29}
\newcommand{\nbznb}{764}
\newcommand{\nbzb}{1151}
\newcommand{\ibznb}{66}
\newcommand{\nbznu}{88}
\newcommand{\nbzu}{227}
\newcommand{\ibznu}{38}

\newcommand{\nsrcc}{2614}
\newcommand{\probtotc}{36.5}
\newcommand{\nbayc}{2529}
\newcommand{\probbayc}{27.9}
\newcommand{\nbayonlyc}{596}
\newcommand{\nlrc}{2018}
\newcommand{\problrc}{130.9}
\newcommand{\nlronlyc}{85}
\newcommand{\nhighlatfc}{591}
\newcommand{\nhighlatblc}{1027}
\newcommand{\nhighlatuc}{941}
\newcommand{\nhighlatagc}{55}
\newcommand{\nhighlatcaa}{540}
\newcommand{\nhighlatcab}{9}
\newcommand{\nhighlatcac}{4}
\newcommand{\nhighlatcad}{38}
\newcommand{\nhighlatcba}{288}
\newcommand{\nhighlatcbb}{270}
\newcommand{\nhighlatcbc}{316}
\newcommand{\nhighlatcbd}{153}
\newcommand{\nhighlatcca}{327}
\newcommand{\nhighlatccb}{128}
\newcommand{\nhighlatccc}{126}
\newcommand{\nhighlatccd}{360}

\newcommand{\nrdgc}{31}
\newcommand{\nagnc}{8}
\newcommand{\nnlsyc}{9}
\newcommand{\ncssc}{5}
\newcommand{\nssrqc}{2}

\newcommand{\nlowlat}{344}
\newcommand{\nlowlatf}{36}
\newcommand{\nlowlatbl}{64}
\newcommand{\nlowlatu}{238}
\newcommand{\nlowlatag}{6}
\newcommand{\nlowlataa}{34}
\newcommand{\nlowlatab}{0}
\newcommand{\nlowlatac}{0}
\newcommand{\nlowlatad}{2}
\newcommand{\nlowlatba}{15}
\newcommand{\nlowlatbb}{7}
\newcommand{\nlowlatbc}{25}
\newcommand{\nlowlatbd}{17}
\newcommand{\nlowlatca}{57}
\newcommand{\nlowlatcb}{9}
\newcommand{\nlowlatcc}{12}
\newcommand{\nlowlatcd}{160}

\newcommand{\nrdgl}{4}
\newcommand{\nagnl}{1}
\newcommand{\nnlsyl}{0}
\newcommand{\ncssl}{0}
\newcommand{\nssrql}{0}

\FPeval{\fracbcu}{round(100*\nhighlatu/\nsrc,1)}
\FPeval{\fracnew}{round(100*(\nsrc/1591)-100,0)}
\FPeval{\fraclat}{round(100*(\nsrc/\nFGLh),0)}
\FPeval{\fracfalse}{round(100*(\probtot/\nsrc),1)}
\FPeval{\nboth}{round(\nbay-\nbayonly,0)}
\FPeval{\fracboth}{round(100*(\nboth/\nsrc),0)}

\shorttitle{Fermi-LAT detected active galactic nuclei}
\shortauthors{The {\em Fermi}-LAT Collaboration}

\begin{document}

\title{The Fourth Catalog of Active Galactic Nuclei Detected by the  {\em Fermi} Large Area Telescope}
\author{
M.~Ajello\altaffilmark{1}, 
R.~Angioni\altaffilmark{2,3}, 
M.~Axelsson\altaffilmark{4,5}, 
J.~Ballet\altaffilmark{6}, 
G.~Barbiellini\altaffilmark{7,8}, 
D.~Bastieri\altaffilmark{9,10}, 
J.~Becerra~Gonzalez\altaffilmark{11}, 
R.~Bellazzini\altaffilmark{12}, 
E.~Bissaldi\altaffilmark{13,14}, 
E.~D.~Bloom\altaffilmark{15}
% ...
}
\altaffiltext{1}{Department of Physics and Astronomy, Clemson University, Kinard Lab of Physics, Clemson, SC 29634-0978, USA}
\altaffiltext{2}{Space Science Data Center - Agenzia Spaziale Italiana, Via del Politecnico, snc, I-00133, Roma, Italy}
\altaffiltext{3}{Istituto Nazionale di Fisica Nucleare, Sezione di Roma ``Tor Vergata", I-00133 Roma, Italy}
\altaffiltext{4}{Department of Physics, Stockholm University, AlbaNova, SE-106 91 Stockholm, Sweden}
\altaffiltext{5}{Department of Physics, KTH Royal Institute of Technology, AlbaNova, SE-106 91 Stockholm, Sweden}
\altaffiltext{6}{AIM, CEA, CNRS, Universit\'e Paris-Saclay, Universit\'e Paris Diderot, Sorbonne Paris Cit\'e, F-91191 Gif-sur-Yvette, France}
\altaffiltext{7}{Istituto Nazionale di Fisica Nucleare, Sezione di Trieste, I-34127 Trieste, Italy}
\altaffiltext{8}{Dipartimento di Fisica, Universit\`a di Trieste, I-34127 Trieste, Italy}
\altaffiltext{9}{Istituto Nazionale di Fisica Nucleare, Sezione di Padova, I-35131 Padova, Italy}
\altaffiltext{10}{Dipartimento di Fisica e Astronomia ``G. Galilei'', Universit\`a di Padova, I-35131 Padova, Italy}
\altaffiltext{11}{Instituto de Astrof\'isica de Canarias, Observatorio del Teide, C/Via Lactea, s/n, E38205, La Laguna, Tenerife, Spain}
\altaffiltext{12}{Istituto Nazionale di Fisica Nucleare, Sezione di Pisa, I-56127 Pisa, Italy}
\altaffiltext{13}{Dipartimento di Fisica "M. Merlin" dell'Universit\`a e del Politecnico di Bari, via Amendola 173, I-70126 Bari, Italy}
\altaffiltext{14}{Istituto Nazionale di Fisica Nucleare, Sezione di Bari, I-70126 Bari, Italy}
\altaffiltext{15}{W. W. Hansen Experimental Physics Laboratory, Kavli Institute for Particle Astrophysics and Cosmology, Department of Physics and SLAC National Accelerator Laboratory, Stanford University, Stanford, CA 94305, USA}
% ...
\begin{abstract}

The fourth catalog  of active galactic nuclei (AGNs) detected by the {\em Fermi Gamma-ray Space Telescope} Large Area Telescope (4LAC)  between 2008 August 4 and 2016 August 2 contains $\nsrc$ objects  located at high Galactic latitudes ($|b|>10\arcdeg$). It includes  85\% more sources than the previous 3LAC catalog based on 4 years of data.   AGNs represent at least \fraclat\% of the high-latitude sources in the fourth {\em Fermi}-Large Area Telescope Source Catalog (4FGL), which covers the energy range from 50 MeV to 1 TeV. In addition,  $\nlowlat$ gamma-ray AGNs  are found at low Galactic latitudes. Most of the 4LAC AGNs are blazars (98\%), while the remainder are other types of AGNs.  The blazar population consists  of   24\%  Flat Spectrum Radio Quasars (FSRQs), 38\%  BL Lac-type objects (BL~Lacs), and 38\% blazar candidates of unknown types (BCUs).  On average, FSRQs display softer spectra and stronger variability in the gamma-ray band than BL~Lacs do, confirming previous findings. All AGNs detected by ground-based atmospheric Cherenkov telescopes are also found in the 4LAC.

\end{abstract}
\keywords{gamma rays: galaxies --- gamma rays: observations --- galaxies: active --- galaxies: jets --- BL Lacertae objects: general --- quasars: general}


\section{Introduction} \label{sec:intro}

Thanks to its broad energy range, excellent sensitivity, and all-sky monitoring capabilities, the {\em Fermi Gamma-ray Space Telescope} Large Area Telescope (LAT) has revolutionized our view of the gamma-ray sky.  The fourth {\em Fermi}-LAT source catalog \citep[4FGL, ][]{4FGL},  based on  the first 8 years of data from the mission,  contains 5064 sources in the energy range 50 MeV to 1 TeV.

The fourth catalog of  AGNs  detected by the LAT (4LAC), presented here, is derived from the 4FGL catalog. 
At high Galactic latitudes, active galactic nuclei (AGNs) represent by far the dominant class of gamma-ray sources in the 4FGL. The vast majority of  these AGNs are of the blazar type, which are characterized by having relativistic jets closely aligned with our line of sight. The  two main classes of blazars are Flat Spectrum Radio Quasars (FSRQs) and  BL Lac-type objects (BL~Lacs), distinguished according to the strength of their optical emission lines. FSRQs have strong, broad emission lines, while BL Lacs have weak, narrow, or no such lines.   In addition to the improvements of the 4FGL relative to previous gamma-ray catalogs, the 4LAC has benefited  from updated methods of associating gamma-ray AGNs with those known at other wavelengths. The 4LAC supersedes the third  catalog of  AGNs  detected by the LAT \citep[3LAC,][]{3LAC}, which was based on 4 years of data.

Gamma-ray AGN catalogs constitute unique resources for a broad range of astrophysics research. Recent applications include: population studies probing the BL Lac-FSRQ dichotomy \citep[e.g. ][]{Ghis17,Nal17}; works on individual sources investigating the connections  between gamma-ray loudness and brightness/polarization  at other observational bands  \citep[e.g. ][]{Ang16, Lic17, Mas17, Fan18, Zar18};  timing  correlations between  activity in the gamma-ray band and other wavelengths \citep[e.g. ][]{Fuh16,Ito16}; and tests of the possible link between gamma-ray AGNs and  sources of ultra high-energy cosmic rays \citep[e.g. ][]{Kag17}  or high-energy neutrinos \citep[e.g. ][]{Pad16,Aar17,Gar19}. These catalogs also enable probes of the extragalactic background light  \citep[EBL, e.g., ][]{Abd18} and the intergalactic magnetic field \citep[e.g., ][]{Bro18,Ack18}, along with a measurement of  the AGN contribution to the extragalactic diffuse gamma-ray background \citep[e.g., ][]{For16,DiM18}.
  
The paper is organized as follows. Section 2 briefly describes  the observations by the LAT and the analysis employed to produce the eight-year catalog. In \S3, we present the methods for associating gamma-ray sources with AGN counterparts and the different schemes for classifying them.  Section 4 describes the contents of the 4LAC fits table and gives the statistics  of the  blazar and non-blazar populations. This section also includes a brief presentation of low-latitude ($|b|<10\arcdeg$) AGNs, which do not formally belong to the 4LAC.  Some of the basic properties of the catalog sources are given in \S5, along with a discussion of the overlap with the AGNs detected at very high energies (VHE; energies above 100 GeV) by atmospheric Cherenkov telescopes. Section 6 summarizes our findings. 

In the following, we use a $\Lambda$CDM cosmology with values from the {\it Planck} results
\citep{Planckcosmo}; in particular, we use $h = 0.67$, $\Omega_m = 0.32$, and
$\Omega_\Lambda = 0.68$, where the Hubble constant $H_0=100h$~km~s$^{-1}$~Mpc$^{-1}$. 
 
\section{\label{sec:obs}Observations with the Large Area Telescope --- Analysis Procedures}

The 4LAC analysis was performed in the context of the 4FGL catalog, which is  briefly summarized here. We refer the reader to the parent paper describing the 4FGL catalog  for details \citep[][]{4FGL}. 
The data were collected over the first 8 years of the mission, from 2008 August 4 (MJD 54682) to 2016 August 2 (MJD 57602). 
The reprocessed  P8R3\_SOURCE\_V2 event class \citep{Bru18} data  were used, with photon energies between 50 MeV and 1 TeV, broadening the energy interval with respect to the 100 MeV -- 300 GeV range of 3FGL \citep{3FGL}.  The increase in acceptance relative to the P7REP class used in 3FGL is 20$\%$  at high energies, accompanied by a better point-spread function (PSF). These improvements are beneficial to the source detection and localization and hence to the counterpart association.  A dedicated diffuse emission model was developed for analyses using the new event class. Weights  penalizing photons with low energies and/or having directions close to the Galactic plane were introduced in the 4FGL likelihood to better account for systematic uncertainties.    
More details are available in Table 2 of the 4FGL paper.  Different spectral models (power-law, log-parabola, power-law with super-exponential cutoff) were tested, and the results are systematically reported in 4FGL. Sources with a maximum likelihood Test Statistic (TS) greater than 25 were retained in 4FGL,  corresponding to a significance just over 4$\sigma$ evaluated for the $\chi^2$ distribution with 4 degrees of freedom. Variability was assessed via both 1-year and 2-month light curves.

\section{Source association and classification}

The associations of 4FGL gamma-ray sources are based on positional coincidence with potential counterparts that display AGN-type spectral characteristics in the radio, infrared, optical, or X-ray bands. 
  A conservative policy adopted early in the mission by the {\em Fermi}-LAT Collaboration is that firm identification is only claimed when correlated variability with a counterpart detected at lower energy has been reported. So far, only 78  AGNs have met this condition (see Table 7 of  4FGL). For the other sources, we use statistical approaches for finding associations between LAT sources and AGNs. The two approaches used here, the Bayesian method and the likelihood-ratio method,   have been extensively described in previous catalogs \citep{LAT10_1FGL,2LAC,3LAC} and are briefly summarized below.   

\subsection{\label{sec:assoc}Source Association}

\subsubsection{\label{sec:gtsrcid}The Bayesian Association Method}

The Bayesian method  was adapted for the {\em Fermi}-LAT catalogs  following the work of \citet{mattox97} developed for the Energetic Gamma Ray Experiment Telescope (EGRET) on the {\em Compton Gamma Ray Observatory}.  This method is described in \cite{LAT10_1FGL} and  implemented with the {\sl gtsrcid} tool\footnote{\url{https://fermi.gsfc.nasa.gov/ssc/data/analysis/scitools/overview.html}}.  The 
angular distance between a LAT source and a candidate counterpart corresponds to the position uncertainty in the case of a real association, while it is driven by  the counterpart density in the case of a false (random) association. The prior distribution is fully characterized by a single number, which is the a-priori probability that a given source of a catalog is the true counterpart of a gamma-ray source. This probability (referred to as the prior in the following) is assumed to be constant for a given catalog and is calibrated via Monte Carlo simulations so that the number of false associations, $N_{\mathrm{false}}$, is equal to the sum of the association-probability complements. For a given counterpart catalog, the prior is found to be close to $N_{\mathrm{assoc}}/N_{\mathrm{tot}}$, where $N_{\mathrm{assoc}}$ is the number of associations from this catalog and $N_{\mathrm{tot}}$ is the number of catalog sources.  A uniform threshold of 0.80 is applied to the posterior probability for the association to be retained. 

The list of catalogs used for the AGN associations with 4FGL sources is given in Table \ref{tab:catalogs}. 
 With respect to 3LAC \citep{3LAC},  updates of counterpart catalogs, e.g., BZCAT \citep{bzcat}, have been used when available. An important addition to the set of catalogs is the Radio Fundamental Catalog\footnote{Available at \url{http://astrogeo.org/rfc}.} \citep[RFC, ][]{Pet19}, with 2720 associations with 4FGL sources (representing 85\% of all AGNs). The high-efficiency of association with VLBI catalogs that are sensitive to parsec-scale emission at 4--8~GHz is attributed to two factors: 1)~ the fact that $\gamma$-ray emission and parsec scale radio emission which originate from contemporary AGN activity are related, 2)~ the scarcity of radio sources with parsec scale emission at 4--8~GHz. The RFC includes many new entries that came from dedicated follow-up observations \citep{Pet13,r:aofus2,r:aofus3} of unassociated gamma-ray sources, triggered by the  publication of  previous {\em Fermi}-LAT catalogs. Applying the Bayesian method to the whole catalog and retaining associations with $P\ge$0.80, the association probability attached to the recent additions (181 sources) are reported as NULL to distinguish them for the others.

\subsubsection{\label{sec:like}The Likelihood-Ratio Association Method}

The Likelihood Ratio \citep[LR, e.g., ][]{Cash79,deRuiter77,Pre83,ss92,Lon98,Mas01,2LAC}  method developed in the {\sl Fermi}-LAT context makes use of large,  relatively uniform surveys in the radio and in X-ray  bands. These surveys  enable us to search for possible counterparts among the faint radio and X-ray sources. The LR method is similar in nature to the Bayesian method but the false association rate is computed from the density of objects brighter than the considered candidate, assessed from the survey log {\rm N}-log {\rm S} distribution. The method for computing the probability that a candidate is the `true' counterpart (called the reliability in this context) is described in detail in \S3.2 of the 3LAC paper \citep{3LAC}. A source is considered as a high-confidence  counterpart of a given gamma-ray source if its reliability  is greater than 0.80 for at least one survey. 

For the LR approach, we analyzed  the  NRAO VLA Sky Survey  \citep[NVSS; ][]{NVSScatalog},  the Sydney University Molonglo Sky Survey \citep[SUMSS; ][]{SUMSScatalog}, the Australia Telescope 20 GHz radio source catalog \citep{AT20G_CAT}, and the {\it ROSAT} All Sky Survey (RASS)  Bright and Faint Source Catalogs \citep{RASSbright,RASS_FAINT_CAT}. We also explored the second RASS catalog \citep{2RXS}, but this attempt  did not lead to further associations.

\subsection{\label{sec:assocresults}Association Results}

The  threshold adopted for the association probability is 0.80 in either method. This value represents a compromise between association efficiency and purity. The fraction of sources associated by both methods is \fracboth \% ($\nboth$/$\nsrc$), with $\nbayonly$  and  $\nlronly$  sources being solely associated with the Bayesian and LR methods, respectively.   The overall false-positive rate is \fracfalse \%, where $N_{\mathrm{false}}$ is calculated  as described in \S \ref{sec:gtsrcid}.   The estimated number of false positives among the $\nnofgl$ sources not previously reported in 3LAC  is $\sprobnew$. 
  
  As in previous LAT AGN catalogs, we define a Clean Sample as those 4LAC sources that did not have any cautionary analysis flags, as described in \S3.7.3 of the 4FGL paper \citep{4FGL}. The most frequent flags are  flag 5 (source close to a brighter neighbor), flag 3
(large flux variation when changing diffuse emission model) and flag 2 (large position shift when changing diffuse emission model).
Table \ref{tab:assoc} compares the performance of the two methods in terms of total number of associations $N_{\mathrm{assoc}}$, estimated number of false associations $N_{\mathrm{false}}$, and number of sources  associated solely via a given method, $N{\mathrm{_S}}$, for the full and Clean samples.

\subsection{\label{sec:classif}Source Classification}

The classification of a source as an AGN primarily relies on its optical spectrum. Other characteristics, like the radio loudness, the presence of a flat/steep radio spectrum between 1.4 GHz and 5 GHz, the broadband emission, the core compactness or the level of radio extended emission, the detection of variability and degree of polarization observed in different bands are used as  ancillary information. If available,  earlier classifications  reported in the literature have been checked.

\subsubsection{\label{sec:optclass}Optical Classification}

The different resources used in the 4LAC for the optical classification are, in decreasing order of precedence:
\begin{itemize}
\item optical spectra from recent intensive follow-up programs  \citep[e.g., ][]{shaw2013,Shaw13,bcu,optcamp6,optcamp5,landoni18,optcamp3,marchesi18,optcamp8,review,refined,optcamp2,sdss1,optcamp1,paiano19,paiano17B,paiano17A,paiano17,optcomp17,ricci15,chi16,deM19}; these data are especially valuable for blazar candidates of previous LAT AGN catalogs that had never been observed.  
\item the optical classification published in the BZCAT list \citep[which is a compilation of sources classified as blazars,][]{bzcat}
\item  spectra available in the literature or from online databases, e.g., the Sloan Digital Sky Survey \citep[SDSS; ][]{sdss0,sdss1}, 6dF Galaxy Survey \citep{6df}, when more recent than the latest version of BZCAT \citep{bzcat}. The latter information was used only if the spectrum was published. 
\end{itemize}
The relevant references are  reported in the electronic table of the catalog. 
We did not use the blazar classes from the Simbad database\footnote{\url{http://simbad.u-strasbg.fr/simbad/}}, since some of them correspond to predictions  based on the {\it WISE}-strip approach \citep{WISE} and were not obtained from spectral observations.

In the 4LAC, we classify the AGN-like gamma-ray detected objects adopting the following terminology:
\begin{itemize}
\item confirmed classifications:
\begin{itemize}
\item FSRQ, BL Lac, radio galaxy, steep-spectrum radio quasar (SSRQ), compact steep spectrum radio source (CSS), Seyfert galaxy, and Narrow-Line Seyfert 1 galaxy (NLSy1) -- these are sources with a well-established classification in the literature and/or an optical spectrum with clear evidence for or lack of emission lines.
\end{itemize}
\item tentative classifications:
\begin{itemize}
\item BCU, blazar candidates of uncertain type: these are considered candidate blazars because the association methods (see \S \ref{sec:gtsrcid} and  \ref{sec:like}) select a candidate counterpart that satisfies at least one of the following conditions:
\begin{itemize}
\item a BZU object (blazars of uncertain/transitional type) in the BZCAT list;
\item a source with multiwavelength data in one or more of the  {\it WISE}, AT20G, RFC \citep{Pet19}, CRATES \citep{CRATES}, PMN-CA \citep{PMNcatalog}, CRATES-Gaps \citep{Hea09}, or CLASS \citep{caccianiga} lists that indicates a flat radio spectrum and shows a typical two-humped, blazar-like spectral energy distribution (SED);
\item a source included in radio and X-ray catalogs not listed above and for which we found a typical two-humped, blazar-like SED \citep[see ][]{Bot07}.
\end{itemize}
The scheme followed in  3LAC whereby BCUs were further divided in three subclasses according to the quality or availability  of their optical spectra has not been reconducted in 4LAC.  The large number of new BCU sources would have made this task excessively manpower intensive. 
    
\item AGN -- for these candidate counterparts the existing data do not allow an unambiguous determination of the AGN type and do not meet any of the criteria to be classified as BCU. Their SEDs display properties typical of radio-loud compact core objects, but the literature information is either incomplete or conflicting for different epochs or wavelengths.
\end{itemize}
At low Galactic latitudes, the surveys include a large number of Galactic sources; therefore the 4FGL class of $|b|<10\arcdeg$ sources associated solely via the LR-method has been  set to the ``unknown'' class as opposed to the ``BCU'' class used by default for sources at larger latitudes. These sources are thus not considered here.

\end{itemize}

\subsubsection{\label{sec:sedclass} Classification based on the broadband spectral energy distribution}

Blazars and more generally radio-loud AGNs can also be classified according to the  peak photon frequency $\nu_\mathrm{s,peak}$ of the synchrotron part of their broadband SEDs. As a large number of 4LAC sources do not have a measured redshift (see Section \ref{sec:z}), the frequency in the observer frame was used. The SEDs  of all 4LAC AGNs were generated using the SED data archive and SED(t)-Builder interactive web-tool 
available at the Italian Space Agency (ASI) Space Science Data Center (SSDC)\footnote{http://tools.ssdc.asi.it/SED/}. Inspection of the error ellipse and the position of the  counterpart was first performed using the Sky Data Explorer at SSDC.   Two different approaches were followed to enable a cross-check. The first was the parametric procedure used in \citet{LAT_SED} and \citet{2LAC}, which is based on the broadband spectral indices $\alpha_{\mathrm{ro}}$ (between 5 GHz and 5000 \AA) and $\alpha_{\mathrm{ox}}$  (between 5000 \AA~and 1 keV).  The list of surveys and catalogs  providing the broadband flux density data is given in \cite{LAT_SED}. The second method, already used in 3LAC and favored here,  consisted of fitting the SED synchrotron  hump with  a third-degree polynomial fit in the log-log plane. This fit was carried out manually on a source-by-source basis after carefully discarding outlying data (e.g., taken during flaring episodes) and those dominated by the thermal emission of the accretion disk or of the host galaxy.  This methodology  allowed us to assign a $\nu_\mathrm{s,peak}$ value to more objects, since a measured X-ray flux is not required, provided the SED curvature is sufficiently pronounced in the IR-optical band. This fit also provided the  $\nu F_{\mathrm{\nu}}$ value at the peak position. Limitations arose from  possible human errors, the use of non-simultaneous broadband synchrotron data, and remaining contamination of thermal emission.  This contamination  may result in an overestimation of the $\nu_\mathrm{s,peak}$ values for  FSRQs, while the near-IR-optical contribution of the host galaxy may bias  $\nu_\mathrm{s,peak}$ low  in BL Lacs. Comparing the results of the two procedures indicated that the fitting method led to an average shift of  $-$0.23 (rms: 0.53) and $-$0.22 (rms: 0.80) in $\log \nu_\mathrm{s,peak}$ relative to the initial method for FSRQs and BL~Lacs, respectively. We identify these shifts as systematic uncertainties. Relative to 3LAC, a more conservative approach was followed when little data were available, leading to a lower fraction of classified sources. The mean differences in  $\log \nu_\mathrm{s,peak}$ between 4LAC and 3LAC are $-$0.14 (rms: 0.37) and $-$0.11 (rms: 0.65) for FSRQs and BL~Lacs, respectively. 

Following \citet{LAT_SED}, the value of the observed $\nu_\mathrm{s,peak}$ was used to classify the source as either a low-synchrotron-peaked blazar (LSP, for sources with $\nu_\mathrm{s,peak} < 10^{14}$~Hz), an intermediate-synchrotron-peaked blazar (ISP, for $10^{14}$~Hz~$< \nu_\mathrm{s,peak} < 10^{15}$~Hz), or a high-synchrotron-peaked blazar (HSP, if $\nu_\mathrm{s,peak} > 10^{15}$~Hz). To obtain the rest-frame value of $\nu_\mathrm{s,peak}$, a correction by a $(1+z)$ factor is needed, where $z$ is the redshift. 


\section{\label{sec:cat} The Fourth LAT AGN Catalog (4LAC)}

\begin{figure}
\centering
\epsscale{2}
\plottwo{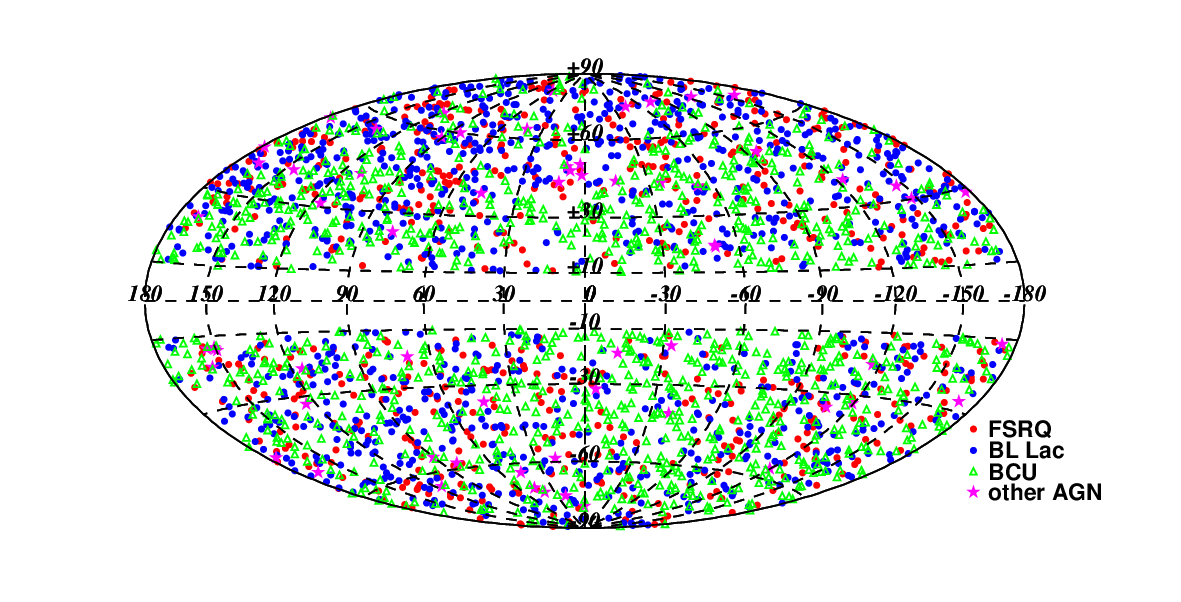}{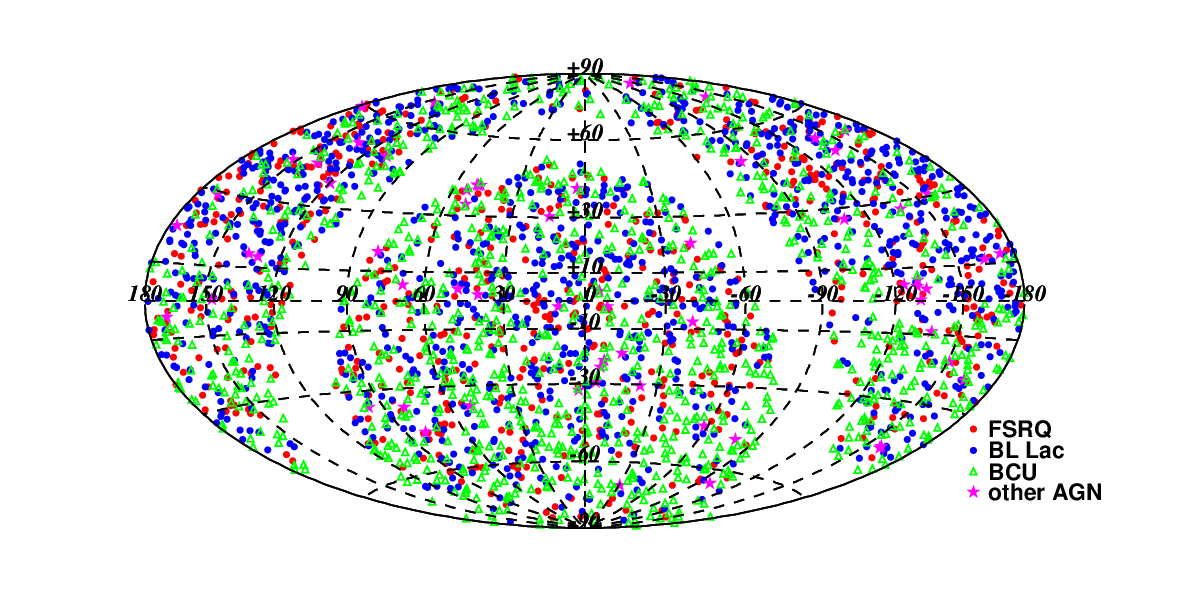}
\caption{Locations of the sources in the Clean Sample in Galactic (top) and  J2000 equatorial (bottom) coordinates and Hammer-Aitoff projection.}
\label{fig:sky_map}
\end{figure}

\begin{figure}
\centering
\resizebox{16cm}{!}{\rotatebox[]{0}{\includegraphics{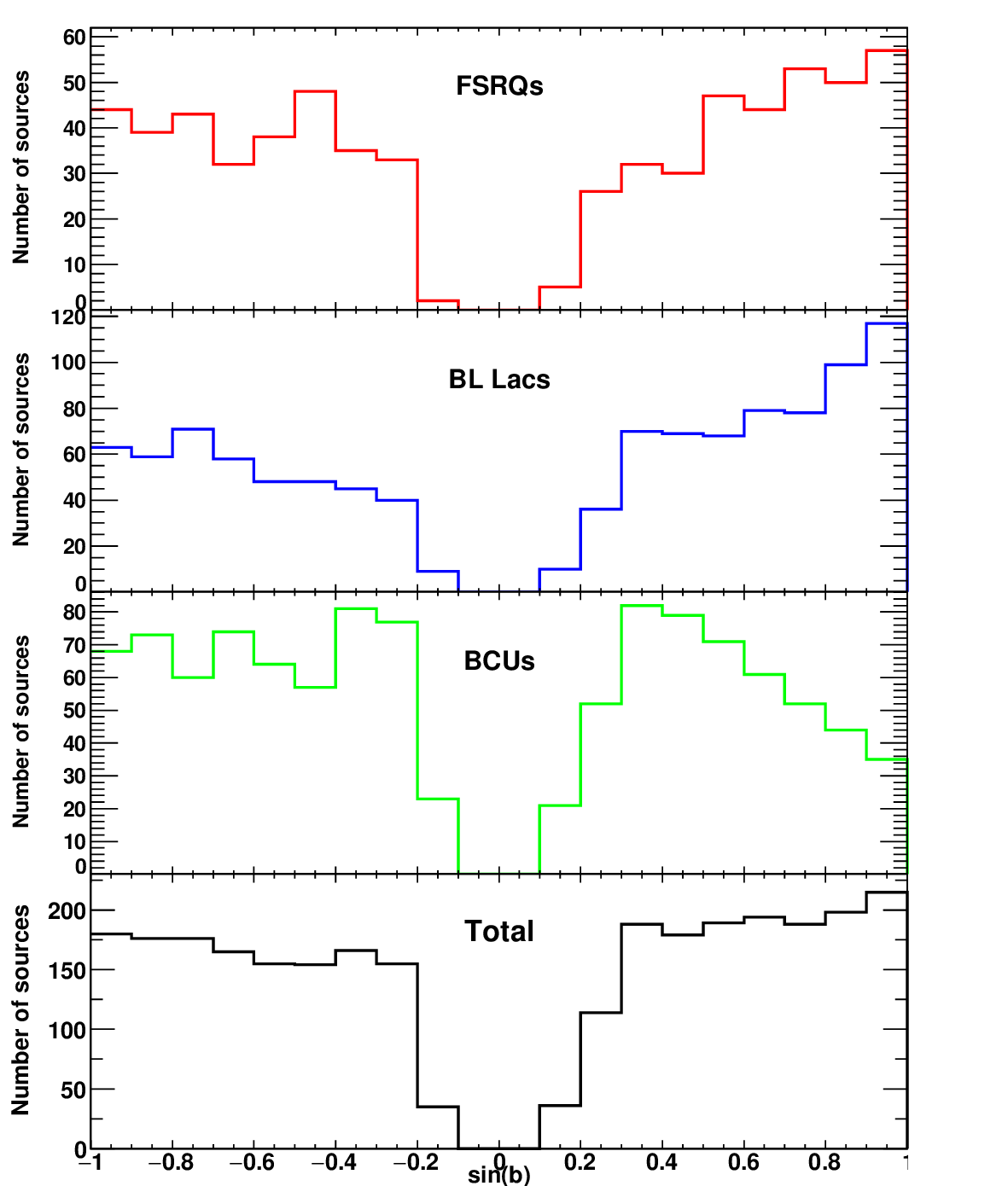}}}
\caption{Distributions of Galactic latitudes for the different blazar classes.}
\label{fig:gal_lat}
\end{figure}

Figure \ref{fig:sky_map} displays the loci of the 4LAC sources in Galactic and J2000 equatorial coordinates. As already noted in previous LAC catalogs, it is clear from this figure that sources of different classes are not uniformly distributed over the sky. This anisotropy is demonstrated in Figure \ref{fig:gal_lat} showing the Galactic-latitude distributions for all 4LAC sources as well as for FSRQs, BL Lacs, and BCUs separately. The anisotropy is most noticeable for BL Lacs, which are 42\% more  abundant in the northern Galactic hemisphere than in the southern one. BCUs show the opposite pattern and somewhat offset the overall anisotropy as seen in the total distribution of sources, which is close to being uniform. The observed anisotropies stem from the larger and better spectroscopic data available in the literature for the northern hemisphere relative to the southern one. Better spectroscopic data are required to assess the BL~Lac nature of an object relative to a FSRQ, because of the weaker optical emission lines in the spectrum of a BL Lac object.   

The format of the 4LAC fits table\footnote{Available at \url{https://www.ssdc.asi.it/fermi4lac/table\_4LAC.fits}.} is described in Table \ref{tab:description}. In addition to relevant parameters\footnote{The FSRQ 3C~454.3 is the only AGN whose preferred spectral shape  is a power law with subexponential cutoff. The corresponding parameters can be found in the 4FGL fits file.} from the 4FGL fits file\footnote{\url{https://fermi.gsfc.nasa.gov/ssc/data/access/lat/8yr\_catalog/gll\_psc\_v22.fit}}, we report the optical and SED-based classes, redshifts, observer-frame synchrotron-peak frequencies, and $\nu F_{\mathrm{\nu}}$ at the synchrotron-peak frequencies. When available, we also provide the VLBI and Gaia counterparts as given in the RFC.    The median position accuracy of VLBI counterparts is 0.8 mas. Therefore, establishing association with VLBI immediately allows us to  propagate the associations to the optical range using Gaia and IR using WISE. Following this route, we obtain Gaia associations with 2134  gamma-ray blazars (74\%).   

\subsection{\label{sec:census} Census}
Table \ref{tab:census} summarizes the 4LAC statistics. The 4LAC includes $\nsrc$  sources, with  $\nhighlatf$ FSRQs,  $\nhighlatbl$ BL~Lacs, $\nhighlatu$ BCUs, and  $\nhighlatag$ other AGNs. A total of $\nnofgl$ sources were not reported in the previous 3LAC catalog, although some of these have been reported elsewhere, e.g., \cite{Ars17,Ars18}. The new sources include  $\nnofglf$  FSRQs,  $\nnofglb$ BL~Lacs,  $\nnofglu$ BCUs, and $\nnagn$ non-blazar AGNs.  The Clean Sample contains  $\nsrcc$  sources, with  $\nhighlatfc$ FSRQs,  $\nhighlatblc$ BL~Lacs, $\nhighlatuc$ BCUs, and  $\nhighlatagc$ other AGNs. The figures shown in the following only include Clean-Sample sources,  unless specified otherwise

\begin{figure}
\centering
\resizebox{16cm}{!}{\rotatebox[]{0}{\includegraphics{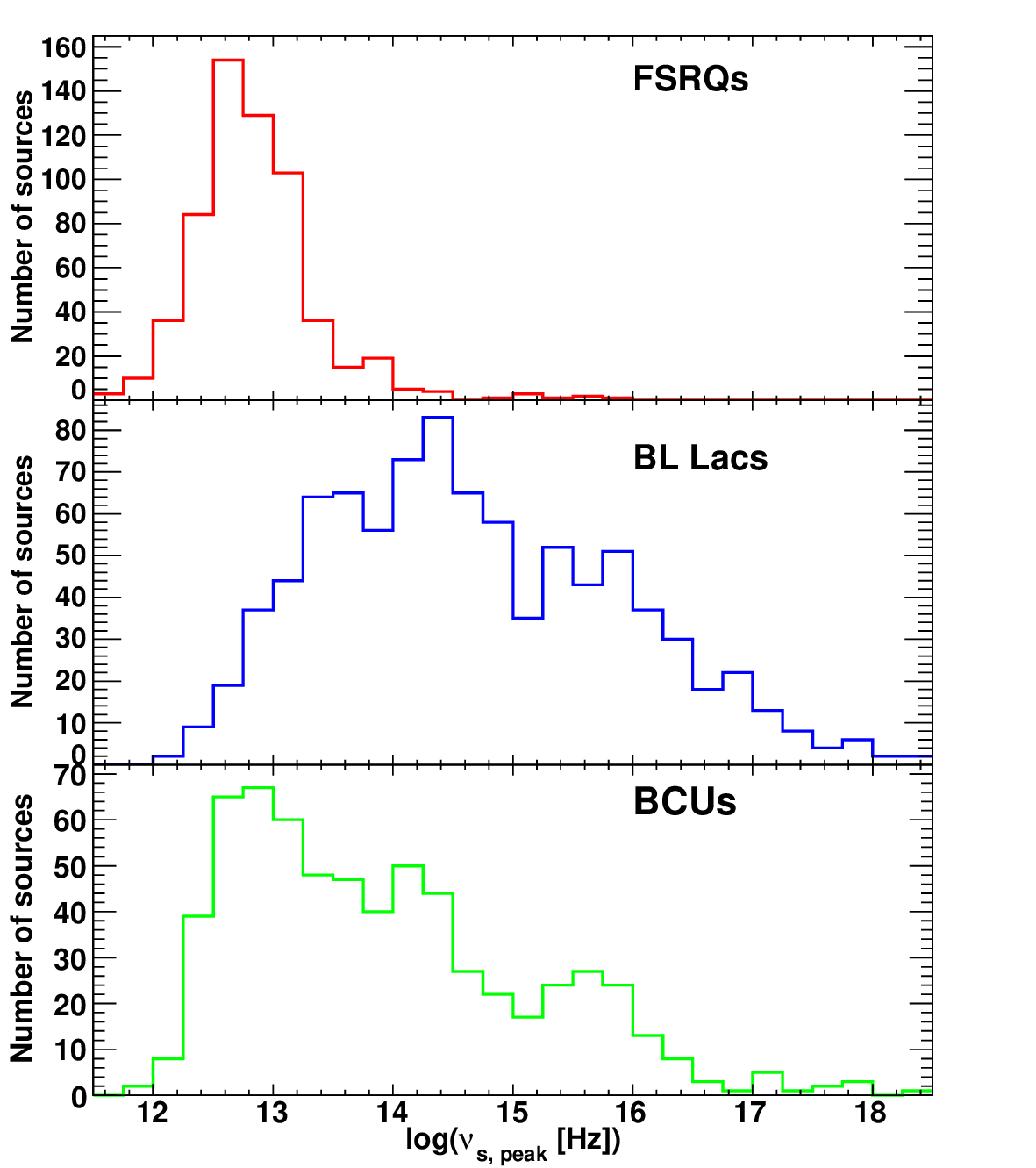}}}
\caption{Distributions of the synchrotron peak frequency $\nu_\mathrm{s,peak}$ for the different blazar classes.} 
\label{fig:syn_hist}
\end{figure}

The SED-based scheme was able to classify  92\% of the FSRQs and 85\% of the BL~Lacs in 4FGL. These values are lower than in 3LAC (99\% and 97\% for FSRQs and BL~Lacs respectively)  due to the more conservative classification procedure used here. The  classified fraction decreases to  60\% for BCUs, for which fewer broad-band archival data are available.  Figure  \ref{fig:syn_hist} shows the  $\nu_\mathrm{s,peak}$ distributions for the different classes. FSRQs are overwhelmingly of the LSP class; therefore no distinction based on SED-based classes is made for them in figures and tallies. In addition to leaving more sources unclassified, the new procedure has produced  a decrease in the the share of ISPs and HSPs among FSRQs, from 10\% to 2\%. Of the five HSP FSRQs, all located at z$<0.63$, two are new:  B3~0038+377 and PKS~1555$-$140.

BL Lacs are fairly evenly distributed among the LSP, ISP and HSP subclasses. The 4LAC  $\nu_\mathrm{s,peak}$ distribution  differs  substantially from the 3LAC one, where HSPs were the most abundant subclass. The new classification procedure caused 15\% of the AGN previously classified as HSP to be reclassified as ISP. 

As a testimony to the substantial follow-up observational efforts already mentioned,  144 sources reported as BCUs in the 3LAC paper (either at high- or low-Galactic latitudes) are now classified as BL~Lacs and 17  as FSRQs. Three BCUs have been reclassified as radio galaxies (IC 1531, TXS~0149+710, PKS~1304$-$215). 
Eight sources have changed from a FSRQ to a BL~Lac (RGB J0250+172, NVSS J040324$-$242946,  GB6~J0941+2721, 2MASS~J11303636+1018245,  PKS~1144$-$379, 4C +15.54, TXS~1951$-$115, PKS~2233$-$173) and three more  from a BL~Lac to  a FSRQ (PMN J0709$-$0255, B2 2234+28A, TXS~2241+406).    

The 3LAC sources that are not present in 4LAC are listed in Table \ref{tab:miss}, along with the various reasons for this situation. Fifty-five 3LAC sources have not been detected in 4FGL. Nineteen 3LAC sources were duplicate associations for which the smaller 4FGL error boxes relative to 3FGL have enabled  the association ambiguity  to be lifted. Fifteen sources reported in 3LAC have lost their associations (becoming unassociated) while three others are  now associated with non-AGN counterparts, all due to improved localizations.

\subsection{\label{sec:misaligned}Non-blazar AGNs and misaligned AGNs}
\FPeval{\totag}{round(\nhighlatag+\nlowlatag,0)}
Table \ref{tab:magn} lists the $\totag$ non-blazar AGNs included in the 4FGL,  $\nhighlatag$ of which belong to the 4LAC and  $\nlowlatag$ are part of the low-latitude sample.  Non-blazar sources represent about 2\% of the total number of AGNs in the 4FGL, a fraction that is basically identical to that found in the 3LAC. Non-blazar sources are further separated into six different classes: 41 radio galaxies\footnote{Two different gamma-ray sources are associated with the core and lobes of Cen A.}, 9 NLSy1s, 5 CSSs, 2 SSRQs, 1 Seyfert galaxy and 11 other AGNs.

A total of $\nnagn$ new non-blazar AGNs are reported in the 4LAC, 22 of which are radio galaxies. The median 1.4 GHz radio luminosity of the newly detected radio galaxies is about $10^{24.4}$~W~Hz$^{-1}$, with the distribution ranging over more than 4 decades (from below $10^{22}$~W~Hz$^{-1}$, for NGC\,2892, to above $10^{26}$~W~Hz$^{-1}$, for PKS~2324$-$02).
The detection of the FR I radio galaxy 3C~120 in gamma rays
was first reported by  \cite{MAGN} using 15 months of
LAT data, but it is reported in a LAT catalog for the first time. Its absence from previous catalogs can be
attributed to periods of flaring interspersed with long periods of low
activity \citep{Tan15}.  TXS~1303+114 and TXS~1516+064 are members of the FRICAT \citep{FRICAT}, although they were earlier proposed as candidate low-power BL Lacs in \cite{Cap15} based on their mid-infrared and optical emission. Among the sources already present in previous LAT catalogs, Fornax A stands out because it is for the first time detected as an extended source \citep[see ][]{Forn16}.   Three  3FGL sources have changed class from BCU to radio galaxy (IC 1531, TXS~0149+710, and PKS~1304$-$215). We associate 4FGL J1346.3$-$6026 with Cen B, although its location is not coincident with that of the radio-galaxy core but points to the southern radio jet. Similarly, we associate 4FGL J1516.5+0015 with the radio galaxy 4C +00.56 (PKS~1514+00, $z=0.0525$) while  the gamma-ray position is closer to the lobes than to the core of the radio galaxy.

The only new SSRQ is 3C~212, for which X-ray emission associated with both lobes was detected by {\it Chandra}  \citep{Ald03}.
Three new CSSs are reported in the 4LAC: 3C~138, 3C~216, and 3C
309.1. These new CSSs are hosted in quasars. At the pc scale they
show a core-jet structure with apparent superluminal motion, indicating
Doppler effects and small viewing angles \citep{Par00,She01,Lis19}.    
While the CSSs all have very high radio luminosity, at the opposite end of the radio luminosity distribution are two other new LAT sources, NGC 3894 and NGC 6328. Based on the small extent of their radio emission, two-sided parsec scale morphology \citep{Tay98,Tin03}, and low radio luminosity, these sources are excellent candidates for being young radio galaxies \citep[see also][]{Mig16,Pri19}.

The classification of a source as a NLSy1 relies on three criteria, as reported in \citet{Ost85}, \citet{Goo89}, and \citet{Pog00}: (i) a full width half maximum (FWHM) of the  H$_\beta$ line $<$ 2000 km s$^{-1}$; (ii) a [OIII]$\lambda$5007/H$_\beta$ ratio$<$3;  and  (iii)  unusually  strong  FeII lines. Nine NLSy1 are reported in the 4LAC. Four of them are new with respect to the 3LAC: 1ERS B1303+515, B3~1441+476, MG2 J164443+2618, and TXS~2116$-$077.  B3~1441+476, MG2 J164443+2618, and TXS~2116$-$077 were previously reported by \citet{DAm15} and \citet{Pal18}, respectively.  The Circinus galaxy \citep{Hay13} remains the only radio-quiet Seyfert galaxy detected by the LAT.

Among other AGNs, there are  two  remarkable cases.  One is PKS~0521$-$36, previously classified as BCU in the 3LAC, which shows a knotty VLBA radio structure similar to misaligned AGN. Based on the broad emission lines in the optical and ultraviolet bands and the steep radio spectrum, a possible classification as an intermediate object between broad-line radio galaxies and SSRQ has been suggested by \citet{Da215}.   The new source PKS~2331$-$240 was the subject of a multiwavelength study revealing features of a giant radio galaxy restarted as a blazar \citep{Her17}.   

Finally, 10 non-blazar AGNs
reported in the 3LAC are not confirmed in the 4LAC. Five of them had a double
association in the 3LAC and are now firmly associated with the other counterpart; two have changed associations  (formerly with 3C~221 and 3C~275.1); one has been reclassified as a BCU (GB 1310+487); while two 3FGL non-blazar AGNs are missing in 4FGL (TXS
0348+013, PKS~1617$-$251).

\subsection{\label{sec:lowlat}Low-Latitude AGNs}

In addition to high-latitude ($|b|>10\arcdeg$) 4LAC sources, we present a low-latitude sample\footnote{The fits file is available at \url{https://www.ssdc.asi.it/fermi4lac/table\_lowlat\_sample.fits}.}. This sample is less complete than the 4LAC because the LAT detection flux limit is higher in this region (by factors of a few) and the counterpart catalogs suffer from Galactic extinction. Because a large contamination of Galactic sources is present in the radio and X-ray surveys used in the LR association method,  the classes of the resulting associations are highly uncertain. Consequently, these associations were not considered here. 
The census of the low-latitude sample is given in the last column of Table  \ref{tab:census}. The fraction of BCUs (62\%) is overwhelming, as expected from the observational hindrance mentioned above.    


\section{Properties of the 4LAC Clean Sample Sources}

\subsection{\label{sec:spectrum} Flux and Spectral Properties}
\begin{figure}
\centering
\resizebox{15cm}{!}{\rotatebox[]{0}{\includegraphics{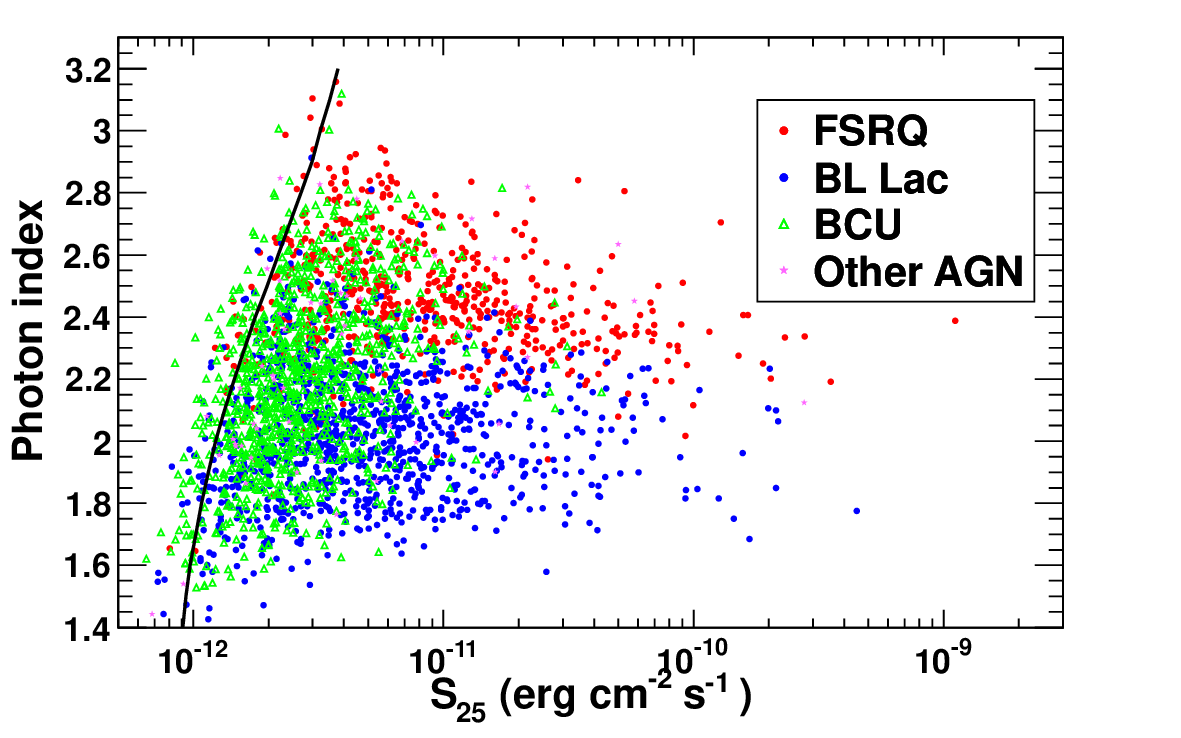}}}
\caption{Photon index as a function of energy flux above 100 MeV. The curve represents the approximate detection limit.  Error bars have been omitted for clarity. The mean photon-index uncertainties are 0.08, 0.10, and 0.14 for FSRQs, BL~Lacs and BCUs respectively.  }
\label{fig:index_S}
\end{figure}

Although many 4FGL/4LAC sources show significant spectral curvature in the gamma-ray band \citep{4FGL}, the gamma-ray power-law photon index, $\Gamma$, represents a convenient way to compare the spectral hardness of different sources across various classes and flux values. This index is plotted vs. the 8-year average energy flux above 100 MeV, $S_{25}$, in Figure \ref{fig:index_S}. The flux detection limit ranges from (1--4)$\times 10^{-12}$ erg cm$^{-2}$ s$^{-1}$ (i.e., close to 1 eV  cm$^{-2}$ s$^{-1}$), with a slight dependence on the spectral hardness. This dependence is stronger than in 3LAC due to the introduction of weights in the 4FGL likelihood to better account for systematic uncertainties. 
BCUs are rare among the brightest sources (i.e., with $S_{25}>$ $10^{-11}$ erg cm$^{-2}$ s$^{-1}$), while they are dominant close to the flux limit.

\begin{figure}
\centering
\resizebox{14cm}{!}{\rotatebox[]{0}{\includegraphics{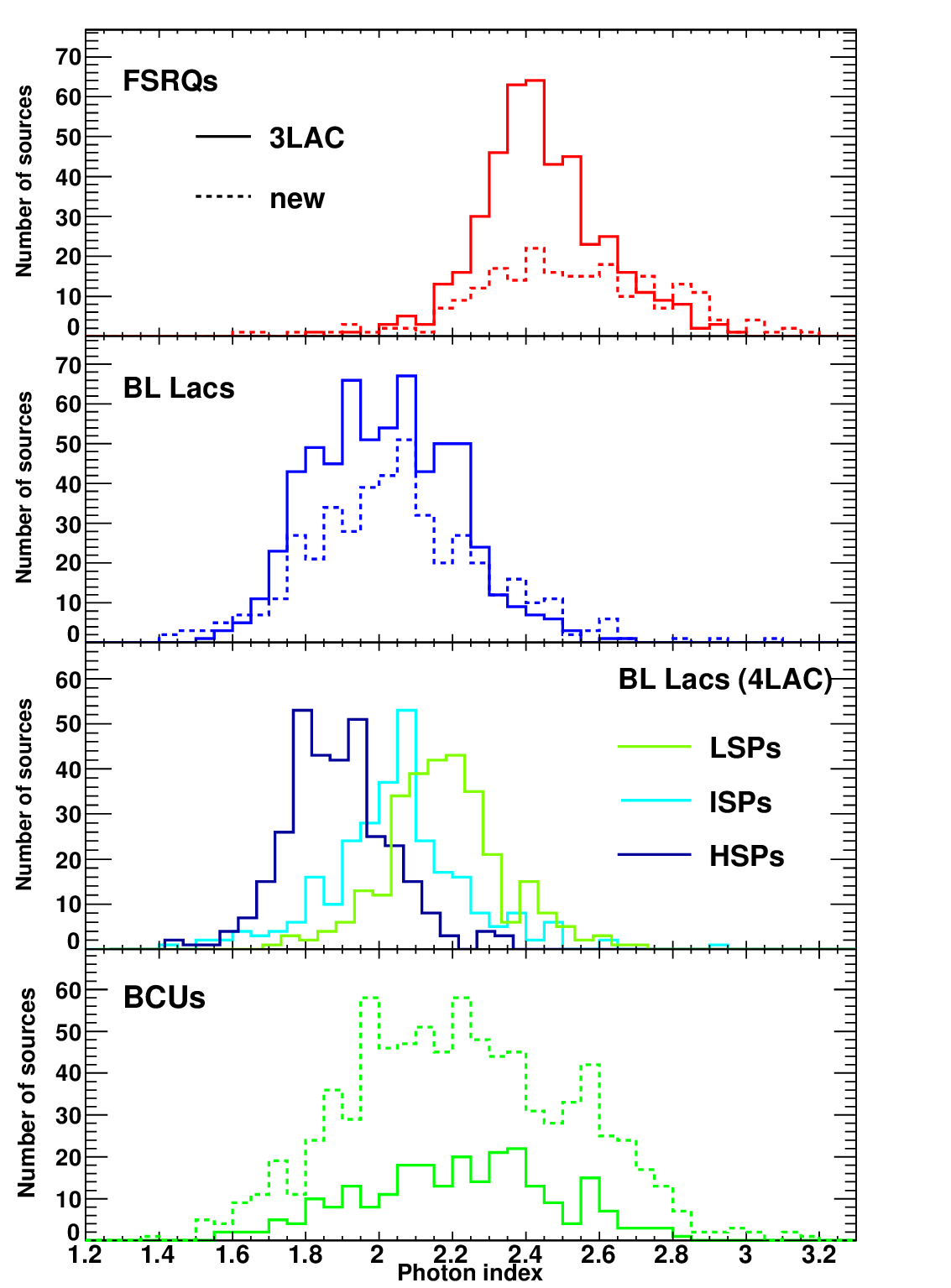}}}
\caption{Photon index distributions for the different blazar classes and subclasses. In the top, second from top and bottom panels, the solid histograms represent the 4LAC sources already present in 3LAC and the dashed histograms the new 4LAC sources.}
\label{fig:index}
\end{figure}

\begin{figure}
\centering
\resizebox{14cm}{!}{\rotatebox[]{0}{\includegraphics{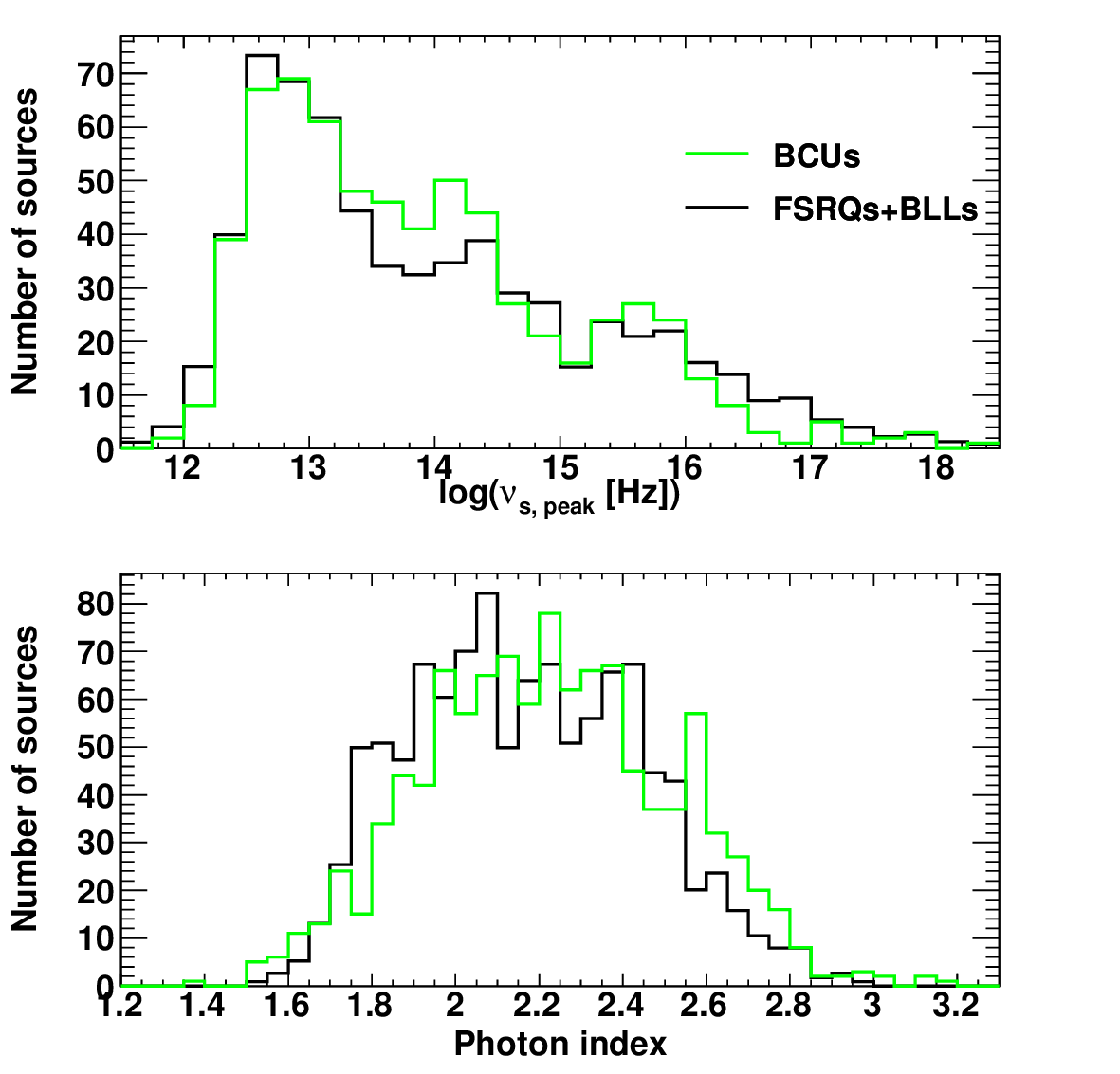}}}
\caption{Comparison between the $\nu_\mathrm{s,peak}$ (top) and photon index (bottom) distributions of BCUs (green) and the (normalized) distributions  obtained by adding up the FSRQ and BL Lac distributions (black). See text for details.}
\label{fig:decomp_bcu}
\end{figure}

Figure \ref{fig:index} displays the photon index distributions for the different optical blazar classes, for 4LAC sources already present in 3LAC and the new ones. Newly detected gamma-ray FSRQs, i.e., not reported in previous LAT AGN catalogs, have somewhat softer spectra  (difference in median $\Gamma$ $\simeq$0.08) than the previously reported ones,  possibly indicating the emergence of sources with SED peaking at lower energy.  The difference between the FSRQ and BL~Lac distributions is striking. The $\Gamma$ medians and rms  are 2.44$\pm$0.20 and  2.02$\pm$0.21 for FSRQs and BL~Lacs respectively. The relative separation in gamma-ray spectral hardness between FSRQs and BL~Lacs already reported in previous LAT catalogs is confirmed: 89\% of FSRQs and 86\% of BL~Lacs have $\Gamma$ greater and lower than 2.25 respectively. This feature by itself carries significant discrimination power between the two classes. The photon index varies among the BL~Lac subclasses, with medians and rms in $\Gamma$ of 2.17$\pm$0.16, 2.05$\pm$0.19, and 1.88$\pm$0.14 for LSPs, ISPs, and HSPs respectively.  The BCU index distribution straddles that of the two classes and extends to $\Gamma$=3. The expectation that both the BCU photon-index and $\nu_\mathrm{s,peak}$ distributions correspond to linear combinations of the observed FSRQ and BL~Lac distributions is tested in Figure \ref{fig:decomp_bcu}. Composite distributions were built assuming the same fractions of FSRQs and BL~Lacs as in the observed sample for the photon index distributions. For the  $\nu_\mathrm{s,peak}$ distribution,  a  slight correction (a factor of 0.92) in the normalization was introduced  in disfavor of FSRQ to account for the difference in efficiency for fitting the SEDs successfully, due to better broad-band data for FSRQs (see section \ref{sec:census}). 
The reasonable agreement between the composite and BCU distributions for  both photon index  and $\nu_\mathrm{s,peak}$ seen in Figure  \ref{fig:decomp_bcu} supports the idea that the sample of unclassified blazars (i.e., BCUs) is of similar composition as the classified sample in 4LAC.  

\begin{figure}
\centering
\resizebox{14cm}{!}{\rotatebox[]{0}{\includegraphics{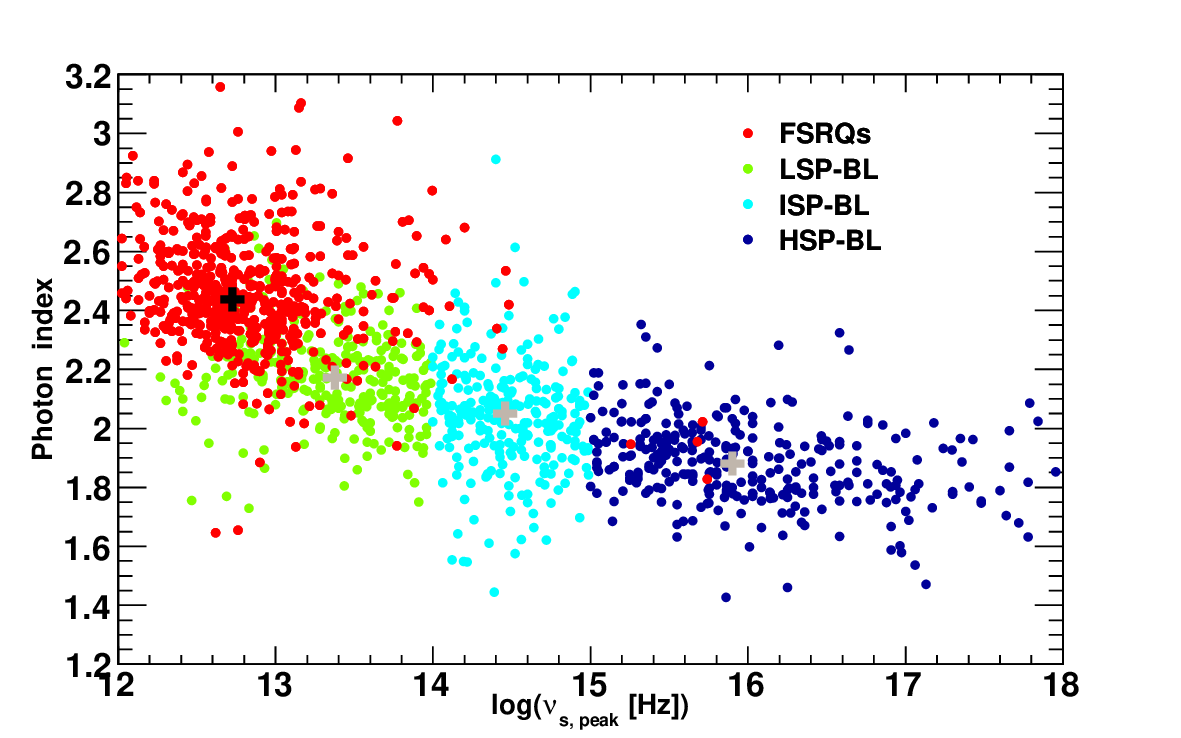}}}
\caption{Photon index vs. frequency of the synchrotron peak $\nu_\mathrm{s,peak}$ in the observer frame. Error bars have been omitted for clarity. The mean photon-index uncertainties are 0.08 and 0.10 for FSRQs and BL~Lacs respectively.  The black cross depicts the FSRQ median photon index, while the gray crosses depict those for the three BL~Lac subclasses.}
\label{fig:index_nu_syn}
\end{figure}

Figure \ref{fig:index_nu_syn} displays the photon index as a function of the observed synchrotron peak frequency.  Even though $\nu_\mathrm{s,peak}$ should be corrected by a factor 1+z to obtain the rest frame value and make the correlation more physical, the fairly strong correlation already noted in previous catalogs is clearly visible. The correlation obtained in 3LAC was reproduced  theoretically by \citet{Der15}  using an equipartition blazar model with a log-parabola description of the electron energy distribution.
In the region of $\nu_\mathrm{s,peak}$  where BL~Lac LSPs and FSRQs overlap, their photon index distributions are very similar. This is the expected  region for  objects that might be transitioning between being FSRQs and BL~Lacs. \citet{Rua14} found 6 of such transitioning objects. Five of them, all of the LSP subclass, are present in 4LAC:  OJ 451 (4FGL J0833.9+4223, FSRQ, $\Gamma=2.44\pm0.07$), TXS~1013+054 (4FGL J1016.0+0512, FSRQ, $\Gamma=2.18\pm0.04$), PKS~1247+025  (4FGL  J1250.6+0217, BLL, $\Gamma=2.00\pm0.10$),  5C 12.291 (4FGL J1308.5+3547, FSRQ, $\Gamma=2.29\pm0.06$), and   PMN J2206$-$0031 (4FGL J2206.8$-$0032,  BLL, $\Gamma=2.25\pm0.05$). Three of these sources have  $\Gamma$ very close to the $\Gamma$=2.25 limit outlined above. 
The four HSP FSRQs in the Clean Sample all have $\Gamma <2.02$, as expected from the class of  "HFSRQs", which  is not  fitting with the "blazar sequence" \citep[see e.g., ][]{pad07}.

Some blazars have undergone statistically significant spectral changes since 3LAC: PKS~1349$-$439 (BL~Lac, $\Delta \Gamma \equiv \Gamma_{4FGL}-\Gamma_{3FGL}=-0.37\pm0.12$), RX J1415.5+4830 (BL~Lac, $\Delta \Gamma=-0.70\pm0.16$), PKS~1532+01 (FSRQ, $\Delta \Gamma=-0.41\pm0.12$) and S4 1800+44 (FSRQ, $\Delta \Gamma=-0.33\pm0.09$). The changes in photon index are all such that $\Gamma_{4FGL}$ is  closer to the median of the class than was $\Gamma_{3FGL}$. Inspection of the light curves reveals that all four sources show enhanced activity in the last 4 years of the 4LAC period relative to that of 3LAC , which may be correlated with the observed spectral hardening. Fourteen  other sources have experienced  spectral variations greater than 3$\sigma$ but of lower amplitudes than these four.  

\begin{figure}
\centering
\resizebox{16cm}{!}{\rotatebox[]{0}{\includegraphics{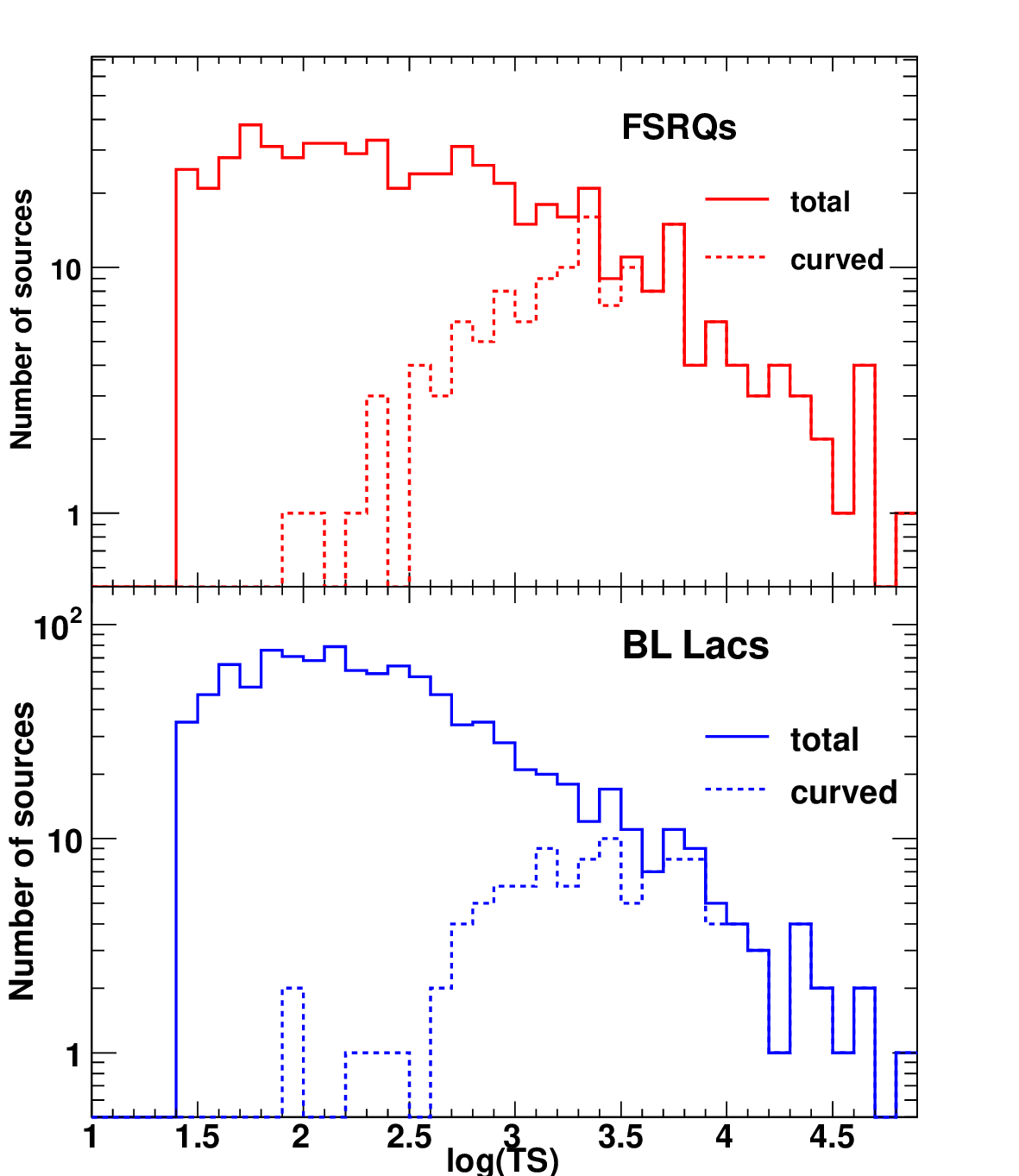}}}
\caption{TS distributions of FSRQs (top) and BL~Lacs (bottom) for curved-spectra sources and the whole sample.}
\label{fig:TS_curv}
\end{figure}

As noted above, many  FSRQs and BL~Lacs show  significant spectral curvature. 
The comparison of the TS distributions of sources with significantly curved spectra to those of the whole sample of FSRQs and BL~Lacs (Figure \ref{fig:TS_curv}) demonstrates that  essentially all bright blazars  have curved spectra in the LAT energy range. A total of 212 FSRQs,  172 BL~Lacs,  and 70 BCUs have  significantly curved spectra.  To enable comparison, if we apply the more stringent threshold on the curvature significance (i.e., approximately 4$\sigma$ instead of 3$\sigma$) used  in  3LAC, these numbers become  146 FSRQs, 112 BL~Lacs, and 26 BCUs. In 3LAC,  the  comparable numbers were 57, 32 and 8, respectively. It is therefore likely that fainter blazars have curved spectra as well, but the current data do not allow their curvature to be established with high confidence.  

\subsection{\label{sec:z}Redshifts}

We conducted a literature search for spectroscopic redshifts. 
Well established redshifts (999) came from BZCAT or the  optical campaigns  mentioned in Section \ref{sec:optclass}.
For the other sources (656), remaining contamination from photometric values or from bad S/N optical spectra cannot be excluded. We found redshifts for all the FSRQs but were unable to find those for 36\% of the BL~Lacs in our sample (compared to 50\% of the BL Lacs without redshifts in 3LAC).  
 This clear improvement 
has been primarily achieved thanks to follow-up observations of 3LAC blazars (see Section \ref{sec:optclass} for references). The fraction without redshifts is similar for the three BL~Lac subclasses (41\%, 42\% and 28\% for LSPs, ISPs, and  HSPs, respectively).

\begin{figure}
\centering
\resizebox{16cm}{!}{\rotatebox[]{0}{\includegraphics{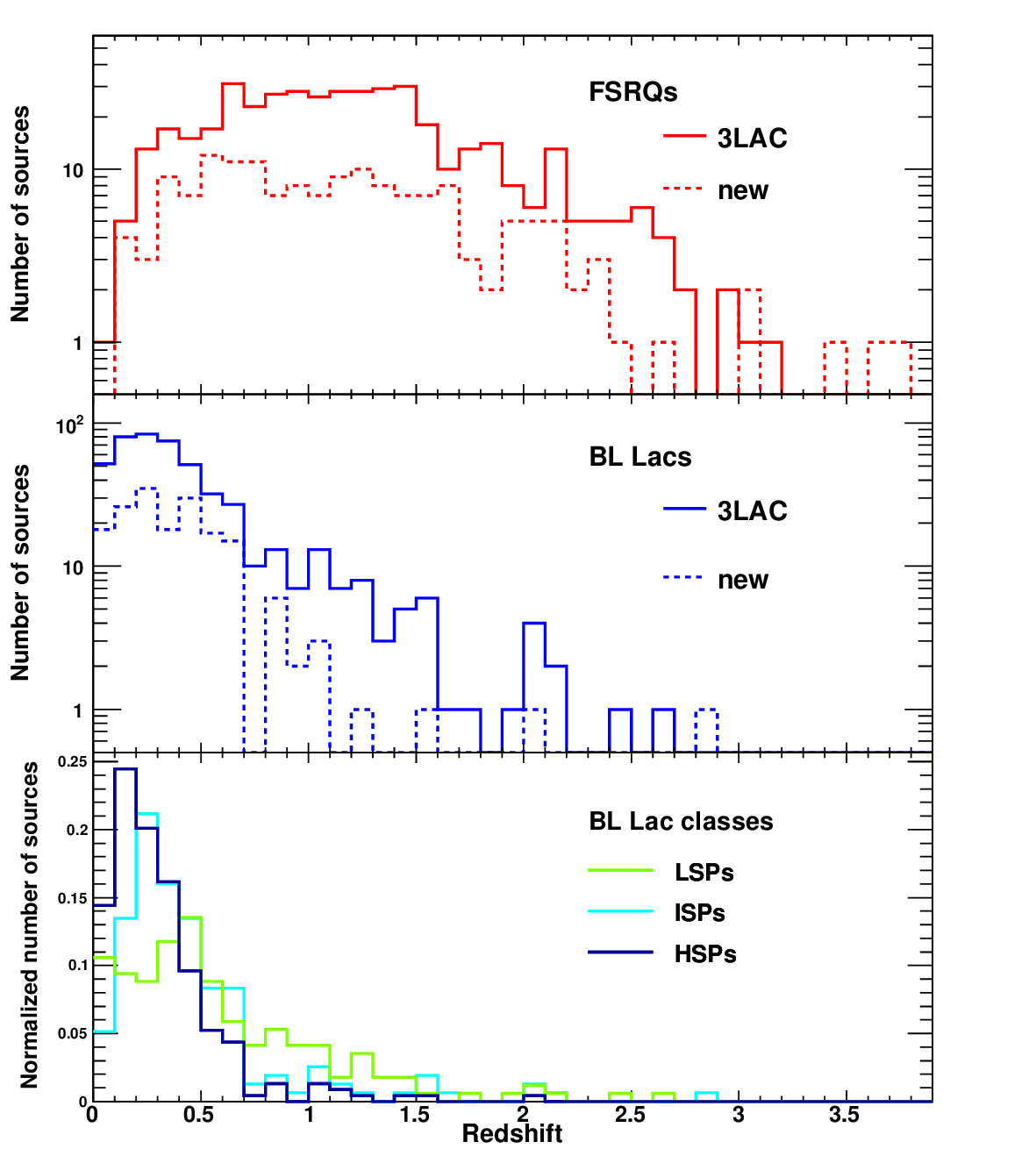}}}
\caption{Redshift distributions (solid: 4LAC sources also in 3LAC, dashed: new 4LAC sources) for FSRQs (top) and  BL~Lacs (bottom)
}
\label{fig:redshift}
\end{figure}

The redshift distributions are displayed in Figure \ref{fig:redshift} for FSRQs and BL~Lacs. The FSRQ distribution shows a broad peak around z=1. This trend confirms the conclusion that the number density of FSRQs grows dramatically up to redshift $\simeq$0.5-2.0 and declines thereafter \citep{Aje12}. For  BL~Lacs, the overall peak lies at z$\simeq$0.3. 
For the sake of comparison, the distributions for  previously and newly reported AGNs are plotted  separately. The redshifts of the 3LAC and newly detected blazars are similar, with  medians and widths of 1.14$\pm$0.62 and 1.09$\pm$0.68  respectively for  FSRQs and  0.34$\pm$0.42 and  0.36$\pm$0.34  respectively for BL~Lacs. The median redshifts decrease between  BL~Lac LSPs, ISPs and  HSPs from 0.47 to 0.36 to 0.25, respectively.
While the maximum redshift for a FSRQ was 3.1 in earlier LAC catalogs, five counterparts to 4LAC sources  have higher redshifts: GB~1508+5714 (z=4.31),  PKS~1351$-$018 (z=3.72), PKS~0335$-$122 (z=3.44), 
MG3 J163554+3629 (z=3.65), and  PMN J0833$-$0454 (z=3.45). GB~1508+5714  and PKS~0335$-$122 are not present in the Clean Sample, however, due to analysis flags. The detections of three of these sources (PKS~1351$-$018,  GB~1508+5714, MG3 J163554+3629  were reported earlier in \citet{Ack17}.  Two other high-z sources  (NVSS J064632+445116 and NVSS J212912$−$153841), whose detections were also  announced in \citet{Ack17}, are absent in the 4FGL/4LAC, possibly due to variability effects.

\begin{figure}
\centering
\resizebox{16cm}{!}{\rotatebox[]{0}{\includegraphics{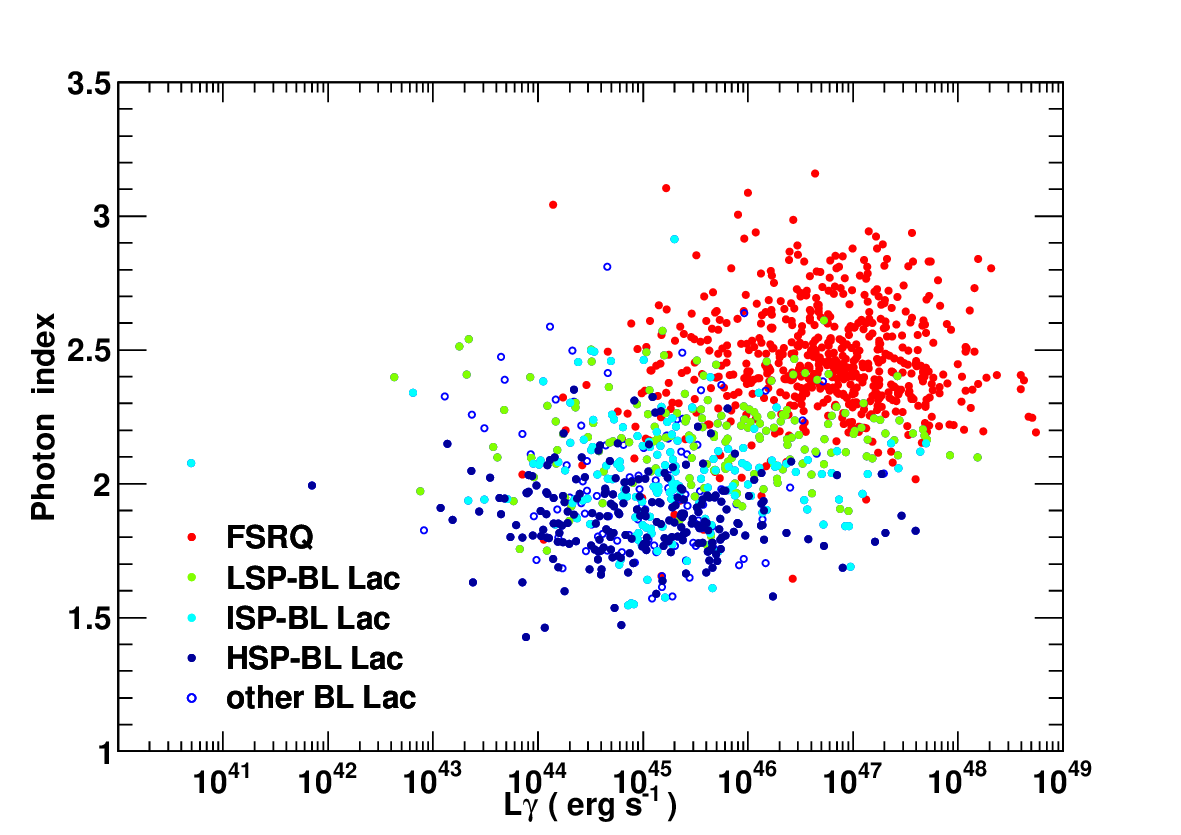}}}
\caption{Photon index vs. gamma-ray luminosity for the different blazar classes and subclasses. Error bars have been omitted for clarity. The mean photon-index uncertainties are 0.08 and 0.10 for FSRQs and BL~Lacs respectively. }
\label{fig:index_L}
\end{figure}

\begin{figure}
\centering
\resizebox{16cm}{!}{\rotatebox[]{0}{\includegraphics{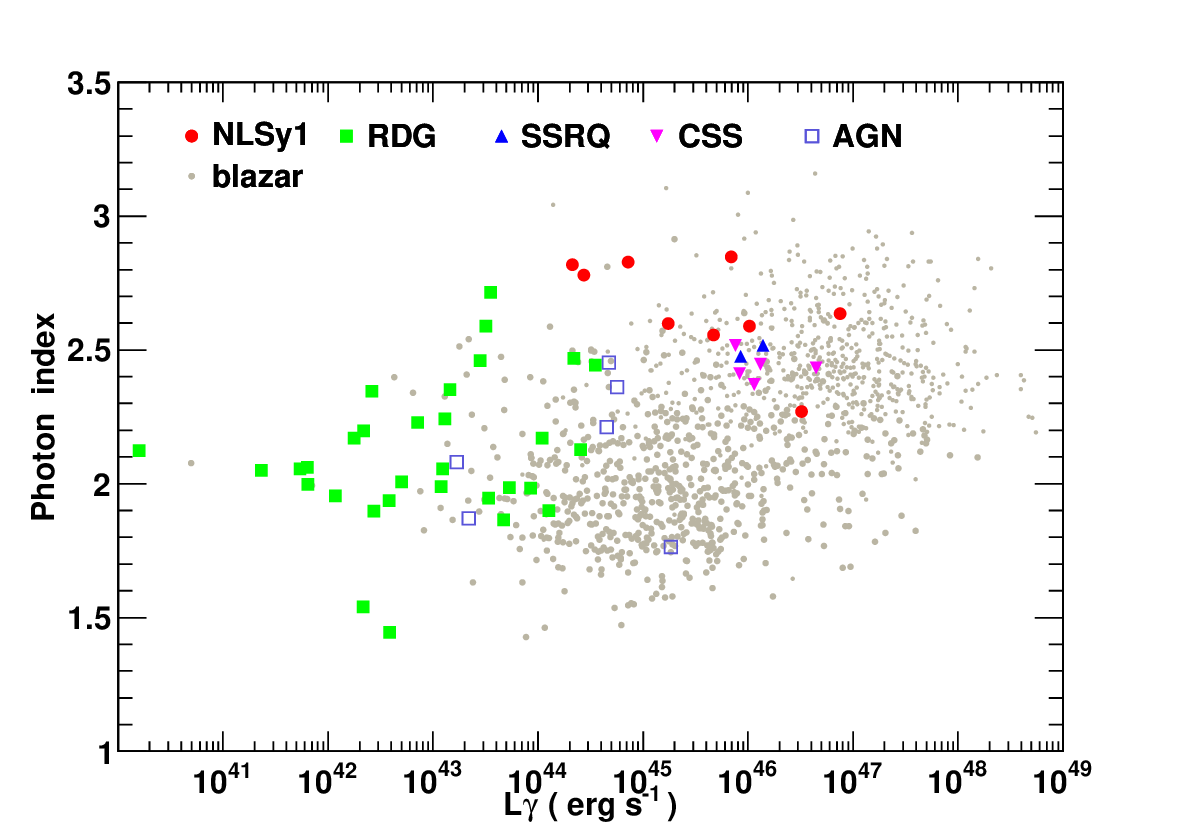}}}
\caption{Photon index vs. gamma-ray luminosity for the different non-blazar classes. Blazars are included for comparison and depicted in gray regardless of their classes. Error bars have been omitted for clarity. The mean photon-index uncertainty is 0.11 for the non-blazar AGNs.  }
\label{fig:index_L_agn}
\end{figure}

Figure \ref{fig:index_L} displays the photon index vs. the gamma-ray luminosity for the different blazar classes. The trend of softer spectra with  higher luminosity observed in earlier catalogs is confirmed. We reiterate the word of caution expressed in 3LAC: this trend is only significant when considering the whole sample of 4LAC blazars. It is not significant when considering the different blazar classes/subclasses individually. Figure \ref{fig:index_L_agn} shows the corresponding plot for the non-blazar sources. Radio galaxies show a large scatter in photon index, while sources of the other classes have fairly soft spectra akin to those of FSRQs. 

\subsection{\label{sec:var} Variability}

Variability is a key property of blazars and is known to depend on the  energy band considered \citep[e.g., ][]{Alek15}. This feature can be naturally explained  as  emitting electrons (assuming a leptonic scenario) of different energies and thus different acceleration/cooling times contribute preferentially to the distinct bands. The assessment of variability in  4FGL does not only depend on intrinsic variability but also on the overall significance of the source detection. Two different values of the variability index\footnote{The variability index is defined as twice the sum of the log(Likelihood) difference between the flux fitted in each time
interval and the average flux over the full catalog interval.} are provided in 4FGL, derived from the sets of 1-year and 2-month light curves. The  1-year light curves allow  the variability of fainter sources to be established compared to the monthly light curves used in early FGL catalogs. Since the variability indices  are distributed as $\chi^2$ functions with $N_{dof}$ degrees of freedom, a   source is defined as variable if at least one of the variability indices is greater than the 99\% confidence limit  of 18.48 and 72.44 for the 1-year ($N_{dof}$=7) and   2-month ($N_{dof}$=47) light curves, respectively.

\begin{figure}
\centering
\resizebox{16cm}{!}{\rotatebox[]{0}{\includegraphics{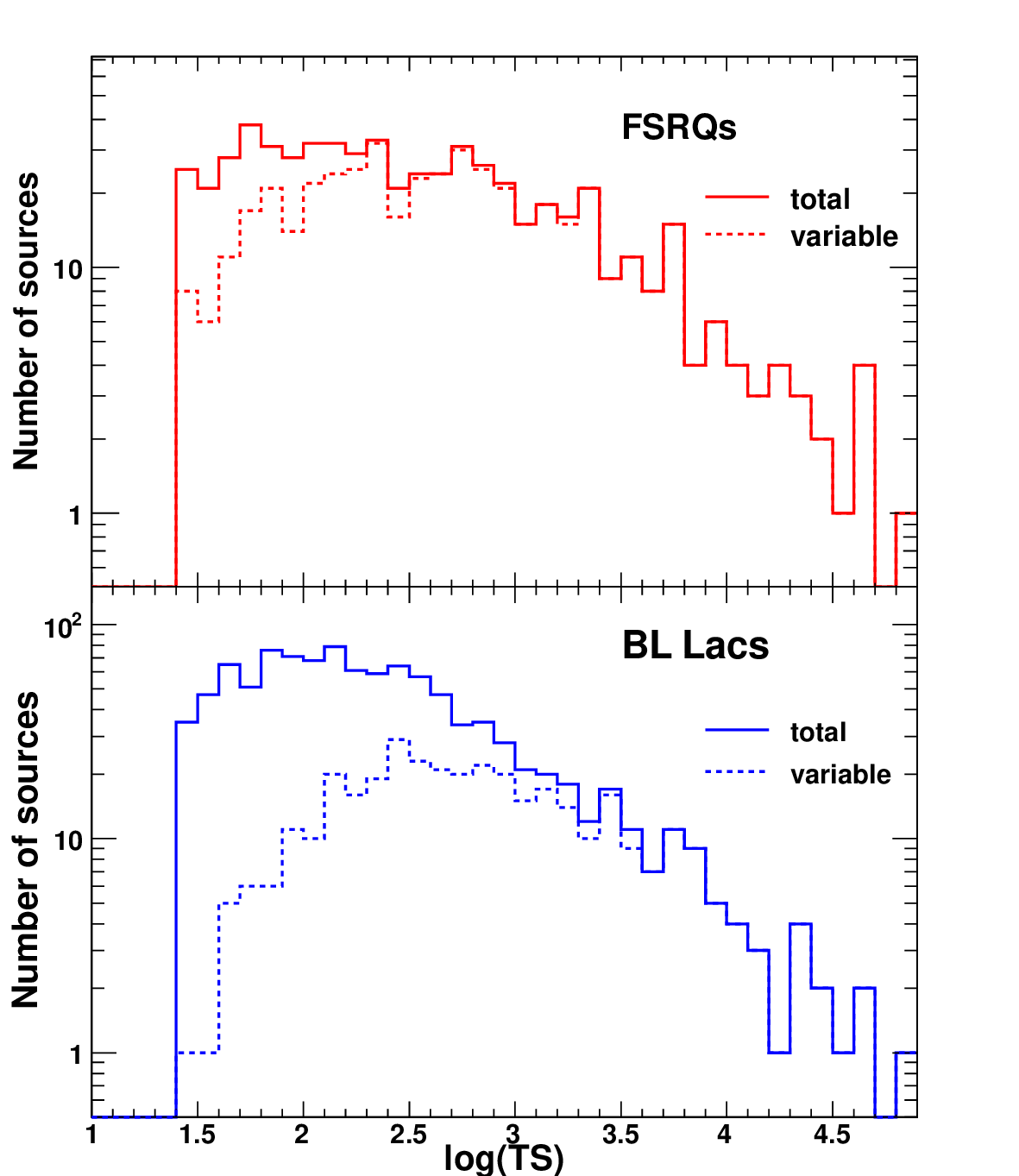}}}
\caption{TS distributions of FSRQs (top) and BL~Lacs (bottom) for variable  sources and the whole sample.}
\label{fig:TS_var}
\end{figure}

Figure \ref{fig:TS_var} displays the TS distributions of variable sources compared to those of the whole sets of FSRQs and BL~Lacs. All bright blazars are found to be variable.
The fraction  
of variable sources goes down from 79\% (464/591) for FSRQs to 35\% for BL~Lacs (362/1027). A monotonic trend is observed for the BL~Lac subclasses, with fractions of 49\% (140/288), 36\% (98/270) and 31\% (98/316) for LSPs, ISPs, and HSPs, respectively. Only  17\% (157/941) of the BCUs show variability, as expected from the fact that these sources tend to be fainter (\S 5.1). The median fractional variability amplitude is 0.63 for FSRQs and 0.27 for BL~Lacs. More extended studies about variability of the LAT-detected blazars can be found in  the 3LAC paper.

Among the radio galaxies, IC~310, NGC~1275, 3C~120, NCG~1218, 3C~111, NGC~2892, and IC~4516 are found to be variable, hinting at a blazar-like behavior for these sources.
 All NLSy1s show significant variability except for IERS B1303+515, B3~1441+476, TXS~2116$-$077. Three out of five CSS sources  are variable (3C~138, 3C~309.1, 3C~380) as well as one  SSRQ (3C~207). The three variable sources of the AGN class are: PKS~0521$-$36, PMN J1118$-$0413, and CGCG 050$-$083.

\subsection{Potential transiently detected AGNs missing in 4LAC}

Some {\it Fermi}-LAT sources show blazar-like flaring activity during limited time intervals  but do not meet the detection significance criterion  to be  included in the 4FGL, which is based on a summed 8-year data set. We present here information about some of these transient sources spatially consistent with AGNs.

When undergoing periods of enhanced activity in  6-hour and 1-day time intervals, some sources can be caught in near-real time by the {\it Fermi}-LAT Flare
Advocate Gamma-ray Sky Watcher (FA-GSW) service \citep[][ and references therein]{Cip12,Tho15}. The brightest sources are often reported to the community in Astronomer's Telegrams (ATels). A total of 371 ATels (plus 3 errata) were posted on behalf of the LAT Collaboration in the 8-year period considered in the 4FGL/4LAC (2008 August 4 to
2016 August 2), and a total of 472 ATels were published from 2008 July 24 (ATel 1628) to 2019 August 15
(ATel 13032). At the time of writing, announced transient sources  that were positionally consistent with blazars or other AGNs include: NVSS~J104516+275136 (ATel 12906); S5~0532+82 (ATel  12902; PKS~2247$-$131 (ATels  9285,
 9620,  11141); 
PKS~1915$-$458 ($z=2.47$,  ATels  2666,  2679 and already reported as missing in the 3FGL/3LAC catalogs);
TXS~0135+291 (ATel  12888); 2MASX~J15441967$-$0649156 ($z\simeq0.04$, ATel   10482); PKS~1251$-$71 (ATel  8215),
PMN~J0508$-$5628 (ATel  6658); PMN~J2010$-$2524 ($z=0.825$, ATel  6553); TXS~1731+152A  (ATel  6395,  6410); PKS~2136$-$642
(ATel  5695); and
PKS~1510$-$319 ($z=1.71$,  ATel  2528).

Sources detected on a 1-week time scale were reported in  the Second {\it Fermi}-LAT All-sky Variability Analysis (FAVA) Catalog \citep[2FAV,
][]{fava}, based on  the first 7.4 years of {\it Fermi}-LAT mission data.   In the FAVA catalog\footnote{https://fermi.gsfc.nasa.gov/ssc/data/access/lat/FAVA/} the analysis was run in 100-800 MeV and 0.8-300 GeV energy bands, 
leading to the identification of  518 flaring gamma-ray sources. Among these sources,  13 were associated
with  established blazars or blazar candidates  that are  included in neither 4LAC or  3LAC: PMN J0231$-$4746, 2MASS J06164292$-$4021527, TXS 0723+220, 4C +38.28 (B2 0913+39), PKS 1200$-$051, PMN
J1322$-$8419, PKS 1354$-$17, RX J1410.5+6100, PKS 1510$-$319, TXS 1534+378, TXS 1731+152A, PKS 1824$-$582, 1RXS J235018.0$-$055928.

For longer timescales, the first catalog of gamma-ray transient sources (1FLT, The Fermi LAT Collaboration in preparation)
reports detections on monthly time intervals during the first 96 months of {\sl Fermi-}LAT operation. This
catalog contains 64 new  gamma-ray sources not present in 4LAC/4FGL or earlier LAT catalogs.  Their mean photon spectral index $\Gamma$ of 2.6 indicates softer spectra than exhibited by the  4LAC sources (mean $\Gamma$ $\sim 2.2$).
These new sources include 24 BCUs (e.g., PKS 1649$-$031, TXS 0209+168, TXS 1601+160, PKS 2108$-$326 and others), 20 FSRQs (e.g., PKS
1524$-$13, TXS 1226+046, PKS 1200$-$051, PMN J2010$-$2524, PKS 1706+006 and others), one BL Lac object (1RXS J112100.6+014515), one
NLSy1 (Mkn 1501), and two radio galaxies (S5 1733+71 and PKS 2236$-$364).

\subsection{Highest-Energy Photons}
\label{sec:hep}

\begin{figure}
\centering
\resizebox{16cm}{!}{\rotatebox[]{0}{\includegraphics{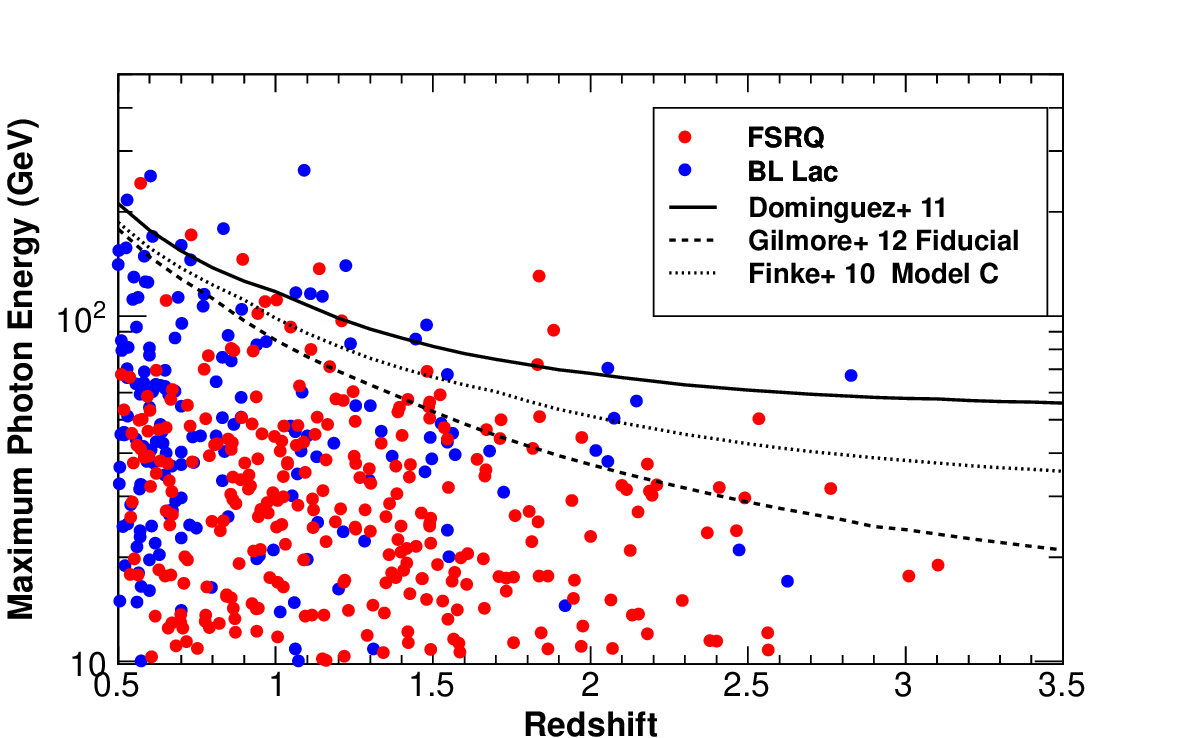}}}
\caption{Energy of the highest-energy photon vs. redshift. The curves display the energies corresponding to an optical depth of 1 as predicted by different models.}
\label{fig:z_he}
\end{figure}

The highest-energy photons detected for each source were selected  within the purest, i.e., with the lowest instrumental background, class (P8R3\_ULTRACLEANVETO\_V2). Based on the energy-dependent PSF, we required a probability\footnote{This probability is derived via the source-to-background ratio defined in appendix A of the 1FGL paper \citep{LAT10_1FGL} and  implemented in the adaptive-binning package available at https://fermi.gsfc.nasa.gov/ssc/data/analysis/user/.} greater than 0.95 for the photons to belong to the source being considered. 
Blazar gamma-ray spectra  provide insight into the density of the EBL via the effect of photon-photon absorption, often defined in terms of an optical depth $\tau$=1, where a photon has only a 1/e probability of reaching the observer \citep[e.g., ][]{Abd18}. A comparison  between the energy of the highest-energy photon measured for a given source with the energy computed for $\tau$=1 at that redshift by theoretical models of the EBL provides a simple and direct test of these models. 
Figure \ref{fig:z_he} compares the 4LAC highest-energy photons with the optical depth predictions of \citet{Fin10},  \citet{Dom11}, and \citet{Gil12}.  Only a few  photons exceed the $\tau$=1 mark, thus remaining compatible with the predictions of these  models.

\subsection{Gamma-Ray Detected versus Non-Detected Blazars}

The  blazars detected in gamma-rays after 8 years of LAT operation represent a sizable fraction of the whole population of known blazars as listed in BZCAT.  This catalog represents an exhaustive list of all sources ever classified as blazars but is by no means complete. Although a comparison between the gamma-ray-detected  and non-detected  blazars within that sample has no strong statistical meaning in terms of relative weights, it is nevertheless useful to look for general trends.

\begin{figure}
\centering
\resizebox{10cm}{!}{\rotatebox[]{0}{\includegraphics{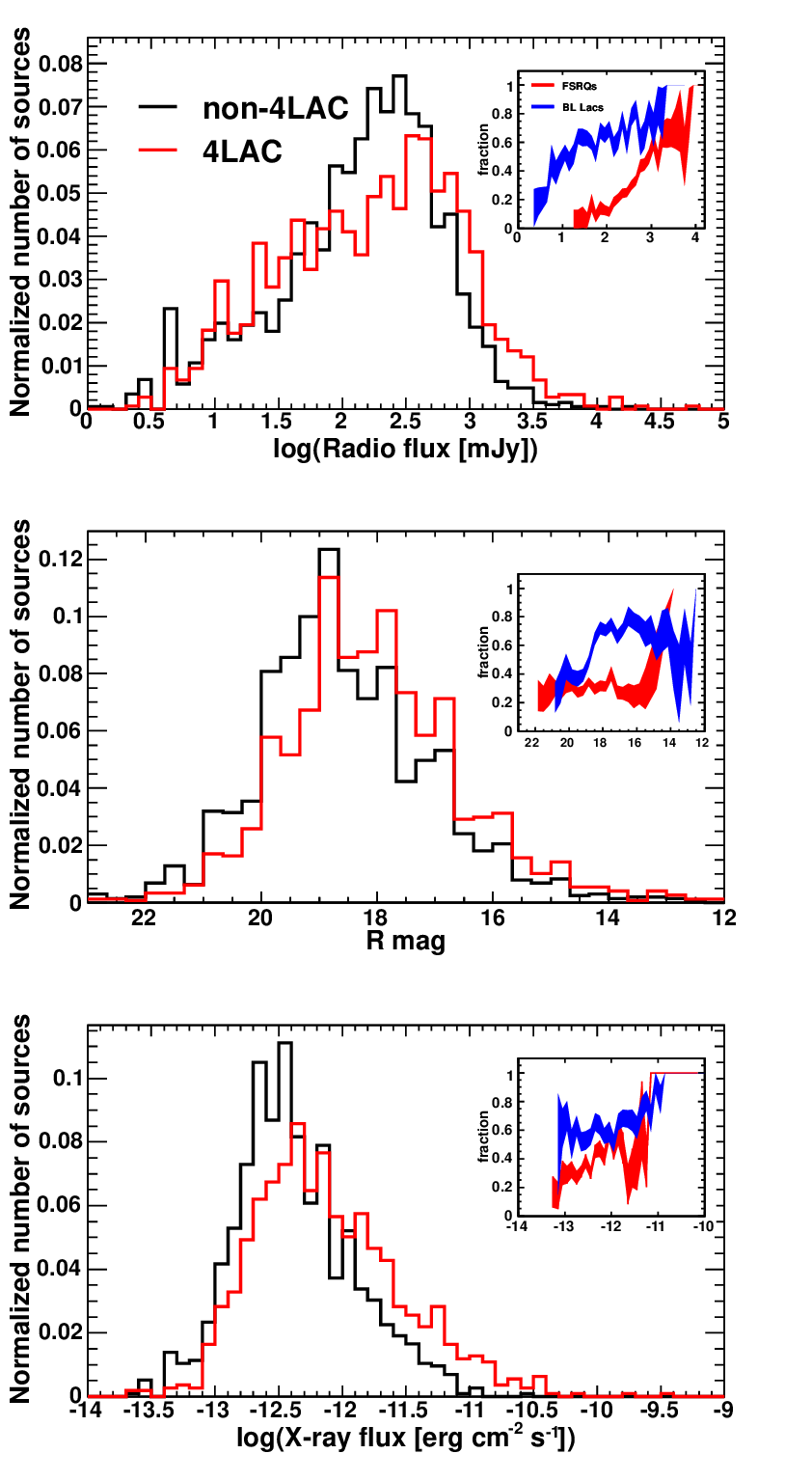}}}
\caption{ From top to bottom: radio flux density at 1.4 GHz, optical R magnitude, X-ray flux (0.1-2.4 keV) distributions for 4LAC (red) and non-4LAC (black) BZCAT sources. The insets show the fraction of 4LAC sources relative to the total for a given flux.  Error bars have been omitted for clarity. }
\label{fig:radio_flux}
\end{figure}

 In total, the 4LAC includes $\ibznf$\% ($\nbznf$/$\nbzf$), $\ibznb$\% ($\nbznb$/$\nbzb$) and $\ibznu$\% ($\nbznu$/$\nbzu$) of the BZCAT \citep{bzcat} FSRQs, BL~Lacs and BCUs respectively. Out of the $\nnofgl$ new 4FGL sources, $\nzz$ are present in BZCAT. Figure \ref{fig:radio_flux} compares the distributions of radio flux at 1.4 GHz, optical  R-band magnitude and  X-ray (0.1-2.4 keV) fluxes for the gamma-ray-detected and undetected BZCAT blazars, as well as the fraction of gamma-ray-detected blazars relative to the total as a function of the different fluxes.   The gamma-ray-detected blazars are somewhat brighter on average in all bands, confirming previous findings \citep{Lis11,Boc16,Pal17}. The fraction of 4LAC blazars in the total population of BZCAT blazars remains non-negligible even at the faint ends of the radio, optical and X-ray flux distributions, in particular for BL~Lacs. This observation is a clue that even the faintest known blazars could eventually shine in gamma rays at LAT-detection levels.      

\subsection{Sources Detected at Very High Energies}
\label{sec:VHE}
Table \ref{tab:GeVTeV} shows the list of 78 AGNs detected by ground-based Cherenkov telescopes, as listed in  TeVCat\footnote{\url{http://tevcat.uchicago.edu/}, as of March 2019}. All are present in 4LAC, and the table shows their optical and SED-based classes, redshifts and 4LAC photon indices. The 3LAC catalog included 55 of the 56 VHE AGNs with detections published or announced at that time. Only  HESS J1943+213 was missing. 
The overall mean  of the 4LAC photon indices for these VHE AGNs is 1.91$\pm$0.20. For the most numerous subclass, the HSP BL~Lacs, we find the mean index to be  1.81$\pm$0.08, i.e., a very hard gamma-ray spectrum. Of the 78, 56 4LAC AGNs are variable at a significance greater than 99\%. 
 
\subsection{\label{sec:indiv} Miscellaneous notes about individual sources}

In the course of the 4LAC analysis, we found a number of individual sources that have changed classification, or have unclear associations.  We include this information about them for completeness.  In case of conflicting associations,  we only  retained the most probable one in the  4LAC catalog. 

\begin{itemize}

\item 4FGL J0140.6$-$0758  is associated with the BL~Lac  RX J0140.7$-$0758. Another object, SDSS J014040.63$-$075857.2, is located 9'' away (i.e., 0.04r$_{95}$, where r$_{95}$ is the 95\% confidence radius),  at  a redshift of 2.674 coming from a  broad-line SDSS optical spectrum. This source has no reported radio emission.

\item  4FGL J0242.3+5216 has two possible high-confidence associations: with GB6 J0242+5209 and  TXS~0239+520, which have similar brightness in the radio band. 

\item  4FGL J0337.8$-$1157 is associated with PKS~0335$-$122, a distant FSRQ at $z=3.442$ that has a damped Ly-$\alpha $ absorption system at $z=3.180$  along the line of sight \citep{Kan03}.

\item 4FGL J0618.1+7819 is associated with the  starburst galaxy  NGC 2146 ($z=0.002975$), but 
the FSRQ 6C 060948+781625 ($z=1.43$) has a similar association probability. 

\item 4FGL J0647.7$-$4418 is associated  with the high-mass X-ray binary RX J0648.0$-$4418 in 4FGL, but the blazar candidate SUMSS J064744$-$441946 has a just barely lower association probability (0.80 vs. 0.85).

\item 4FGL J0720.0$-$6237 is associated with the FSRQ PMN J0716$-$6240,  but  another FSRQ, PMN J0719$-$6218, lies just outside the 95\% confidence region and might contribute to the gamma-ray emission.

\item 4FGL J0814.4+2941 is associated with the BL~Lac  EXO 0811.2+2949, but a broad-line quasar,  SDSS J081425.89+294115.6, also lies  within the  95\% confidence region.

\item  4FGL J0941.9+2724 may have a double association with the BL~Lac 5BZBJ0941+2722 and the  FSRQ MG2 J094148+2728, which was an association in 3FGL.

\item 4FGL~J1300.4+1416  (3FGL~J1300.2+1416) is associated with OW~197 (PKS~1257+145, $z=1.1085$), as it was in 3LAC. At about 5' distance from OW~197, a blazar candidate NVSS~J130041+141728  lies in both the 3FGL and 4FGL 95\% confidence ellipses, despite a  reduction in size of the ellipse by a factor of 4. The 4FGL source position is about midway between OW~197 and NVSS~J130041+141728.

\item 4FGL J1625.7+4134  may have a double association, with 4C +41.32 and  B3~1624+414.

\end{itemize}

\section{Summary}

The 4LAC, derived from the 4FGL catalog, based on 8 years of {\sl Fermi}-LAT data,  includes  $\nnofgl$  (85\%) more AGNs than the 3LAC. At high-Galactic latitudes, AGNs represent at least  $\fraclat$\% of the 4FGL sources. Unassociated sources lying in this sky region share common spectral features with BCUs (see Figure 22 of the 4FGL paper), suggesting that most of them are AGNs as well.  BL~Lacs and FSRQs represent 38\% and 24\% of the blazar population respectively. The increase of the fraction of BCUs in the sample from 29\% in 3LAC to 38\% in 4LAC emphasizes the value of the spectroscopic endeavor carried out by several groups.  From their photon index and  $\nu_\mathrm{s,peak}$ distributions, BCUs probably contain similar fractions of (still unclassified) FSRQs and BL~Lacs as observed in the classified population. 

Fits of the synchrotron-peak positions have been performed manually for all 4LAC sources, leading to a SED-based classification for 75\% of the 4LAC blazars. The number of non-blazar AGNs has almost doubled relative to 3LAC, from 32 to $\nhighlatag$, including 22 new radio galaxies.     
The overall properties of the 4LAC AGNs are similar to those found in 3LAC. A fairly clear separation in spectral hardness between BL~Lacs and FSRQs is observed, with a transition around $\Gamma$=2.25. Five HSP FSRQs, which are not fitting with the "blazar sequence" picture,  are present in 4LAC. They all show  spectra harder than average for the FSRQ class.  Significant spectral curvature is observed for essentially all the brightest (TS$>$3000) blazars.       The redshift distributions of newly detected blazars resemble those found for 3LAC. Five new FSRQs have redshifts greater than the highest 3LAC  redshift (z=3.1), reaching z=4.31. The correlation between photon index and gamma-ray luminosity is strong overall for blazars, but much weaker if the different classes are taken separately.  Analysis of 1-year and 2-month light curves shows that 79\% of the FSRQs and 35\% of the BL~Lacs are variable, along with 7 radio galaxies and 16 other AGNs. The highest-energy photons show  compatibility  with the EBL $\gamma-\gamma$ attenuation predicted by some recent models.  About 30\% of the new blazars are present in BZCAT. Although the 4LAC blazars are predominantly associated with higher-than-average  radio, optical and X-ray fluxes in BZCAT, they remain non-negligible even at the faint ends of these flux  distributions, in particular for BL~Lacs. All 78 known VHE blazars are detected by the LAT, with 56 of them being variable in the GeV range.

\section{Acknowledgments}

\acknowledgments The \textit{Fermi} LAT Collaboration acknowledges generous
ongoing support from a number of agencies and institutes that have supported
both the development and the operation of the LAT as well as scientific data
analysis.  These include the National Aeronautics and Space Administration and
the Department of Energy in the United States, the Commissariat \`a l'Energie
Atomique and the Centre National de la Recherche Scientifique / Institut
National de Physique Nucl\'eaire et de Physique des Particules in France, the
Agenzia Spaziale Italiana and the Istituto Nazionale di Fisica Nucleare in
Italy, the Ministry of Education, Culture, Sports, Science and Technology
(MEXT), High Energy Accelerator Research Organization (KEK) and Japan
Aerospace Exploration Agency (JAXA) in Japan, and the K.~A.~Wallenberg
Foundation, the Swedish Research Council and the Swedish National Space Board
in Sweden. Additional support for science analysis during the operations phase is gratefully acknowledged from the Istituto Nazionale di Astrofisica in Italy and the Centre National d'\'Etudes Spatiales in France.
This work performed in part under DOE Contract DE- AC02-76SF00515.

This research has made use of data obtained from the high-energy Astrophysics Science Archive
Research Center (HEASARC) provided by NASA's Goddard
Space Flight Center; the SIMBAD database operated at CDS,
Strasbourg, France; the NASA/IPAC Extragalactic Database
(NED) operated by the Jet Propulsion Laboratory, California
Institute of Technology, under contract with the National Aeronautics and Space Administration. This research has made use of data archives, catalogs and software tools from the ASDC, a facility managed by the Italian Space Agency (ASI).
Part of this work is based on the NVSS (NRAO VLA Sky Survey).
The National Radio Astronomy Observatory is operated by Associated Universities, Inc., under contract with the National Science Foundation.
This publication makes use of data products from the Two Micron All Sky Survey, which is a joint project of the University of
Massachusetts and the Infrared Processing and Analysis Center/California Institute of Technology, funded by the National Aeronautics
and Space Administration and the National Science Foundation.
This publication makes use of data products from the Wide-field Infrared Survey Explorer, which is a joint project of the University of California, Los Angeles, and
the Jet Propulsion Laboratory/California Institute of Technology,
funded by the National Aeronautics and Space Administration.
Funding for the SDSS and SDSS-II has been provided by the Alfred P. Sloan Foundation,
the Participating Institutions, the National Science Foundation, the U.S. Department of Energy,
the National Aeronautics and Space Administration, the Japanese Monbukagakusho,
the Max Planck Society, and the Higher Education Funding Council for England.
The SDSS Web Site is http://www.sdss.org/.
The SDSS is managed by the Astrophysical Research Consortium for the Participating Institutions.
The Participating Institutions are the American Museum of Natural History,
Astrophysical Institute Potsdam, University of Basel, University of Cambridge,
Case Western Reserve University, University of Chicago, Drexel University,
Fermilab, the Institute for Advanced Study, the Japan Participation Group,
Johns Hopkins University, the Joint Institute for Nuclear Astrophysics,
the Kavli Institute for Particle Astrophysics and Cosmology, the Korean Scientist Group,
the Chinese Academy of Sciences (LAMOST), Los Alamos National Laboratory,
the Max-Planck-Institute for Astronomy (MPIA), the Max-Planck-Institute for Astrophysics (MPA),
New Mexico State University, Ohio State University, University of Pittsburgh,
University of Portsmouth, Princeton University, the United States Naval Observatory,
and the University of Washington.

\bibliography{AGN_4LAC.bib}

\begin{appendix}

%
\section{Chronological convention for source association naming}

In 3LAC and 4LAC, an approximate chronological scheme is adopted for the proper names of the radio/IR/optical/X-ray counterparts. For convenience and more clarity, a time-ordered list of the source catalogs used here is given in Table A1. The proper names follow approximately the initial (discovery) names from radio, IR, optical, and X-ray surveys, catalogs and observations. These names have been checked to be recognized by the  strict ``name resolver'' in the NASA-IPAC Extragalactic Database (and therefore by many other web databases like HEASARC derived from it).

Radio galaxies, quasars, blazars and other AGNs were first discovered as optical non-starlike nebulae objects (i.e. galaxies) listed in  the C. Messier catalog in year 1791, NGC (J.L.E. Dreyer) and IC catalogs published between the years 1781 and 1905.  Examples of sources in 4LAC are the radio galaxies M 87, NGC 315, NGC 1275 (also known as Per A or 3C 84), and the starburst galaxies M 82, NGC 253, and NGC 1068. Other AGNs were initially considered as optical variable stars (Argelander, e.g., BL Lac, W Com, AP Lib and BW Tau, known as 3C 120). Some blazars and AGNs were discovered as unusual optically blue starlike objects (Ton, PHL, Mkn catalogs, about 1957-1974, for example Ton 599, Mkn 180, Mkn 421, Mkn 501, PHL 1389) or in optical galaxy catalogs (CGCG, MCG, CGPG, ARP, UGC, Ark, Zw/I-V, Tol, during the period 1961-1976 (for example CGCG 050-083, UGC 773, I Zw 187, V Zw 326). Subsequent optically-selected objects and quasar catalogs provide some  names of 4LAC associations  (for example, PG, PB, US, SBS, PGC, LEDA, HS, SDSS).

In parallel the largest fraction of radio galaxies and AGNs were discovered during the early era of radio astronomy, with objects like Vir A, Cen A, Cen B, Per A, appearing already in the first half of the 1950s, and the well-known point radio source catalogs 3C, CTA, PKS, 4C, O[+letter], VRO, NRAO, AO, DA, B2, GC, S1/S2/S3, all published between about 1959 and 1974. The PKS (Parkes Radio Catalog, Australia) is the source name preferred for southern celestial radio AGNs, while northern  radio AGNs were likely first reported in Cambridge catalogs (especially 3C, 4C) and the Ohio State University Radio Survey Catalog (Ohio Big Ear radio-antenna, O(x) catalog prefix), or in the CTA, NRAO, DA, B2, TXS, S1-5, or MG1-4 catalogs. Examples in the 4LAC are Cen A (already known as NGC 5128, but better known with its original radio name), Per A, Vir A (M 87). Other radio catalogs published between about 1974 and the mid 1980s are: TXS, 5C, S4/S5, MRC, B3, while from the end of the 1980s until the end of 1990s we have MG1/MG2/MG4, 87GB, 6C/7C, JVAS, PMN, EF, CJ2, FIRST, Cul, GB6, FBQS, WN, NVSS, CLASS, IERS, SUMSS, CRATES. Some catalogs at IR or UV frequencies are also considered here (KUV, EUVE, 2MASSi, 2MASS). 
Further blazars and other AGNs, fainter in the radio band, were discovered directly thanks to the observations made by the first X-ray satellites (2A, 4U, XRS, EXO, H/1H, MS, 1E, 1ES, 2E, RX catalogs published from about 1978 to the mid '90s). Later came the  {\it ROSAT} survey catalogs from re-analysis and cross-matches such as RGB, RBS, RHS, 1RXS, XSS.

Table A1 reports catalog and survey prefixes in an approximate chronological order, adopted for the associations names of the 4LAC catalog. The order is only approximate due to the lack of precise information for each catalog and by the need to follow, in some cases,  the criterion of the most-used name in the literature and published papers for a source, even if this latter choice is rather arbitrary and subject to opinion.

The most frequent source association names in 4LAC come from the 3C, 4C, PKS, O[+letter], B2, S2/S3/S5, TXS, MG1/MG2, PMN, GB6, SDSS, 1ES, RX, RBS, 1RXS, NVSS and 2MASS catalogs.
Some 4LAC counterpart proper names may be somewhat inadequate  because of the radio extension of the AGN/radio galaxy. In some cases the gamma-ray position may relate to the radio emission of the jet or lobe of an AGN, while the name refers to the radio core, which could be offset by a few arcsec.  This is manifest in the two different gamma-ray point-source components of Cen A (designed as Cen A core and Cen A lobe in the 4FGL). Another example is 4FGL J1758.7$-$1621 associated with AT20G J175841$-$161703 (also known as NVSS J175841$-$161705). This steep-spectrum AT20G radio source has a brighter neighbor (23.9 Jy at 160 MHz), PMN J1758$-$1616 (not included in the 3C catalog because it lies 4 degrees off the Galactic plane) at about 10 arcsec offset. The radio structure map reveals that  PMN J1758$-$1616 is a FR-II radio galaxy,  while AT20G J175841$-$161703 corresponds to its radio lobe.

\setcounter{table}{0}
\renewcommand{\thetable}{A\arabic{table}}
\begin{deluxetable}{llcl}
\label{tab:names} 
\tabletypesize{\scriptsize}
%
\tablecaption{Historical catalogs reference for naming of 4LAC associations}
%
\tablehead{\colhead{\normalsize{Waveband}} & \colhead{\normalsize{Publ. year(s)/range}} & \colhead{\normalsize{Prefix }} & \colhead{\normalsize{Catalog name}} 
}
\startdata
Optical & 1781 & M & Charles Messier catalog of nebulae and non-star-like objects from M 1 to M 110.\\
Optical & 1844 - $\sim$ 1915 & xy+const. & Argelander convention for first-discovered variable stars in each constellations. \\
Optical & 1848 & GC & General Catalog  \\
Optical & 1888 & NGC & New General Catalog. \\
Optical & 1896, 1905 & IC & IC Index Catalogs (IC I and IC II, expansions of the NGC Catalog). \\
Radio & 1947-1949 & const.+letter & Constellation + arabic letter (first radio sources ever discovered). \\
Optical & 1952, 1963 & PLX & Yale General Catalog of Trigonometric Stellar Parallaxes \\
Radio & 1955 & 2C & Second Cambridge Radio Catalog at 178 MHz \\
Optical & 1957-1959 & Ton & Tonantzintla (Mexico) Catalog of Blue Stars \\
Radio & 1959-1962 & 3C & Third Cambridge Radio Catalog at 178 MHz (3C, 3CR) \\
Radio & 1960 & CTA & Caltech Radio Survey List A \\
Optical & 1961-1968 & CGCG &  Catalog of Galaxies and of Clusters of Galaxies \\
Optical & 1962 & PHL & Palomar-Haro-Luyten Blue Stellar Objects list \\
Optical & 1962-1974 & MCG & Morphological Catalog of Galaxies \\
Radio & 1964-1967 & Kes &  Kesteven catalog of galactic radio sources \\
Radio & 1964-1968, 1971-1975 & PKS &  Parkes catalog of radio sources \\
Radio & 1965-1969 & 4C & Fourth Cambridge Radio Catalog \\
Radio & 1965-1971 & O(x) (Ohio (x)) & Big Ear Ohio State University Radio Survey Catalog (O + R.A.hour letter) \\
Radio & 1965-1971 & VRO &  Vermillion Radio Observatory survey catalog \\
Radio & 1966 & NRAO & National Radio Astron. Obs. Positions and Flux Densities of Radio Sources \\
Optical & 1966 & ARP & Arp Peculiar Galaxies catalog \\
Radio & 1967-1970 & AO &  Arecibo Occultation Radio Sources \\
Optical & 1967-1974 & Mkn (Mrk) & Markaryan blue object list (Galaxies with an ultraviolet continuum) \\
Radio & 1968 & DA &  Dominion Radio Observatory Survey, List A \\
Radio & 1970-1974 & B2 &  Second Bologna Catalog of radio sources \\
Radio & 1971-1972 & GC &  Green Bank Radio Survey List C \\
Radio & 1971-1972 & S1/S2/S3 &  First/Second/Third "Strong" (radio) Source survey \\
Radio & 1971-1972 & GB (GB1) & Green Bank Radio Survey \\
Radio & 1971-1978 & GB2 &  Green Bank Radio Survey 2 \\
Optical & 1971  & CGPG & Catalog of Selected Compact Galaxies and of Post-Eruptive Galaxies \\
Optical & 1973 & UGC & Uppsala General Catalog of Galaxies \\
Radio & 1974-1983 & TXS &  Texas Survey of Radio Sources \\
Optical & 1974 & UGCA & Uppsala General Catalog Appendix \\
Radio & 1975 & 5C & Fifth Cambridge Survey of Radio Sources \\
Optical & 1975 & Ark & Arakelian Emission Line Objects \\
Optical & 1975 & I-V Zw & First/Second/Third/Fourth/Fifth Zwicky list of compact galaxies \\
Optical & 1976 & Tol & Tololo List of Emission Line Galaxies \\
Optical & 1976, 1983, 1986 & PG & Palomar-Green Bright Quasar Catalog \\
Optical & 1977-1984 & PB & Palomar-Berger Faint Blue Stars Catalog \\
Radio & 1978 & S4 & Fourth "Strong" (radio) Source survey \\
X-ray & 1978 & 2A & Second ARIEL V survey catalog \\
X-ray & 1978 & 4U & Fourth Uhuru Catalog of X-ray Sources \\
Radio & 1978-1995 & GRA (GR) & Grakovo Radio Decametric Survey  \\
X-ray & 1979 & XRS & X-Ray Source catalog from rockets, balloons, satellites of 1964-1977 \\
Ultraviolet & 1980-1984 & KUV & Kiso Ultraviolet Excess Objects catalogs \\
Optical & 1980-1993 & USNO (IDS) & U.S. Naval Observatory parallaxes catalog \\
Radio & 1981 & S5 & Fifth "Strong" (radio) Source survey \\
X-ray & 1981 & 3A & Third ARIEL V survey catalog \\
Gamma-ray & 1981 & 2CG & Second COS-B catalog of high-energy gamma-ray sources \\
Optical & 1981-1984 & US & Usher Faint Blue Stars \\
Radio & 1981, 1991 & MRC & Molonglo Reference Catalog of Radio Sources \\
Radio & 1981, 1994 & 1Jy & Extragalactic radio sources with flux densities $>$1Jy at 5 GHz catalog \\
Optical & 1983-2000 & SBS & Second Byurakan Survey of Emission Line Objects \\
X-ray & 1983-1986 & EXO & EXOSAT X-Ray Source Catalog \\
X-ray & 1984 & H (1H) & The HEAO A-1 X-Ray Source Catalog \\
X-ray & 1984 & EXO & EXOSAT XRay Source Catalog \\
Radio & 1985 & B3 & Third Bologna Catalog of radio sources \\
Radio & 1985-1993 & 6C & Sixth Cambridge Radio Catalog \\
Radio & 1986 & MG1 & First MIT-Green Bank 5 GHz Survey \\
Optical & 1987 & AM & Arp and Madore Southern Peculiar Galaxies and Associations catalog   \\
Infrared & 1988 & IRAS & Infrared Astronomy Satellite Point Source Catalog \\
Infrared & 1990 & IRAS F & 	Infrared Astronomy Satellite Faint Source Catalog \\
Optical & 1989 & PGC & Principal Galaxy Catalog \\
Optical & 1989 & LEDA & Lyon-Meudon Extragalactic Database catalog \\
Optical & 1989 & [HB89] & Hewitt and Burbidge QSO compilation (mute prefix) \\
Optical & 1989-1996 & CTS & Calan-Tololo Survey of galaxies and quasars \\
Radio & 1990 & MG2 & Second MIT-Green Bank 5 GHz Survey \\
Radio & 1990 & MG3 & Third MIT-Green Bank 5 GHz Survey \\
Optical & 1990 & GSC & Hubble Guide Star Catalog \\
Radio & 1990-1998 & 7C & Seventh Cambridge Survey of Radio Sources  \\
X-ray & 1990 & 1E & The Einstein Observatory (HEAO 2) catalog of IPC X ray sources. \\
Radio & 1991 & MG4 & Fourth MIT-Green Bank 5 GHz Survey \\
Radio & 1991  & 87GB & The 1987 Green Bank Radio Survey \\
Optical & 1991-2007 & HS (HE) & Hamburg/ESO QSO Survey \\
X-ray & 1991 & MS & Einstein (HEAO-2) Medium Sensitivity Survey \\
X-ray & 1991 & 1ES & 1st Einstein (HEAO-2) Slew Survey Source Catalog \\
Radio & 1992  & ZS & Zelenchuk Survey \\
Gamma-ray & 1992-2000 & GRO & Compton Gamma Ray Observatory source  \\
Radio & 1992-2002 & JVAS & Jodrell Bank-VLA Astrometric Survey \\
Optical & 1993  & HIP & Hipparcos Catalog \\
Radio & 1994 & PMN & Parkes-MIT-NRAO Radio Survey catalog \\
Radio & 1994 & EF & Effelsberg Radio Sources catalog \\
Radio & 1994 & CJ2 & Second Caltech-Jodrell Bank VLBI Survey catalog \\
Ultraviolet & 1994 & EUVE & Extreme UltraViolet Explorer Bright Source List \\
X-ray & 1994 & RX & First ROSAT Source Catalog of pointed observations with the PSPC \\
X-ray & 1994 & 2E & Second Einstein (HEAO-2) Observatory catalog of IPC X-ray sources \\
Radio & 1995-1997 & FIRST & Faint Images of the Radio Sky at Twenty Centimeters \\
Radio & 1995 & Cul & Culgoora Radio Sources catalog \\
X-ray & 1995 & 1WGA & First White Giommi Angelini ROSAT X-Ray sources list\\
Radio & 1996 & GB6 & Green Bank 6cm Radio Survey \\
Radio & 1996 & CJF & Caltech-Jodrell bank Flat spectrum survey \\
Ultraviolet & 1996 & 2EUVE & 2nd Extreme Ultraviolet Explorer source catalog \\
Radio & 1996-2001 & FBQS & FIRST Bright QSO Survey  \\
Radio & 1997, 2008 & WN & WENSS North radio survey \\
X-ray & 1997 & RGB & ROSAT-Green Bank source catalog \\
X-ray & 1997 & EXSS	& Einstein (HEAO-2) Extended X-Ray Sources \\
Radio & 1998 & NVSS & NRAO VLA Sky Survey \\
Radio & 1998-2002 & CLASS & Cosmic Lens All-Sky Survey catalog  \\
Radio & 1998-2010 & IERS &  International Earth Rotation Service  \\
Radio & 1998-2010 & ICRF &  International Celestial Reference Frame  \\
Infrared & 1998-2000 & 2MASSi & 2 Micron All Sky Survey point sources Incremental release \\
X-ray & 1998-2000 & RBS & ROSAT Bright Survey catalog \\
X-ray & 1998-2000 & RHS & ROSAT Hard X-ray Spectra source catalog \\
Gamma-ray & 1999 &  3EG	&  Third EGRET Catalog of High-Energy Gamma-Ray Sources \\
Optical & 1999 &  MRSS &	Muenster Red Sky Survey \\
Optical & 1999-2008 & SDSS & Sloan Digital Sky Survey Catalogs \\
X-ray & 1999-2000, 2006, 2009 & 1RXS & ROSAT All-Sky Survey Bright Source Catalog \\
Radio & 2000 & VSOP & VLBI Space Observatory Programme \\
X-ray & 2001, 2005 & 1AXG  & 1st ASCA X-ray survey from GIS  experiment \\
Radio & 2003, 2008 & SUMSS & Sydney University Molonglo Sky Survey catalog  \\
Radio & 2003-2009 & WMAP & WMAP Foreground Source Catalogs  \\
Infrared & 2003-2006 & 2MASS &  2 Micron All Sky Survey Point objects Final Release \\
X-ray & 2004 & XSS & RXTE XTE Slew Survey catalog \\
X-ray & 2004-2008 & IGR	& INTEGRAL Gamma Ray source \\
Ultraviolet & 2005-2015  & GALEXASC & GALaxy Evolution eXplorer All-Sky Survey Source Catalog \\
X-ray & 2005 & SHBL & Sedentary High energy peaked BL Lacs \\
X-ray & 2005-2015 & SWIFT & Swift source list  \\
Gamma-ray &  2006-... & HESS &  High Energy Stereoscopic System observatory source list \\
Radio & 2007 & CRATES & Combined Radio All-Sky Targeted Eight GHz Survey  \\
Radio & 2007 & VIPS & VLBA Imaging and Polarimetry Survey \\
Radio & 2007 & VERA & VLBI Exploration of Radio Astrometry \\
Radio & 2007-2008 & VLSS & VLA Low-frequency Sky Survey \\
X-ray & 2007 & SAXWFC & Beppo-SAX X-Ray Satellite Wide Field Camera catalog \\
X-ray & 2007-2010 & 1XMM & 1st XMM-Newton Serendipitous Source Catalog \\
X-ray & 2007-2010 & 2XMM & 2nd XMM-Newton Serendipitous Source Catalog \\
Radio & 2008 & CGRaBS & Candidate Gamma-Ray Blazar Survey  \\
Gamma-ray & 2008 & EGR & Revised catalog of gamma-ray sources detected by EGRET  \\
Radio & 2008-2010 & AT20G & Australia Telescope 20-GHz Survey catalog  \\
Gamma-ray & 2008-... & VER & VERITAS gamma-ray source list \\
Gamma-ray & 2009 & 1AGL	& First AGILE GRID Catalog of High Confidence Gamma-Ray Sources \\
Multifrequency & 2009-2018 & BZ(x) (2-5BZ(x)) & Roma Blazar catalog (the last published has prefix 5BZ+(letter)) \\
Infrared & 2010  & AKARI-IRC-V1 & AKARI/IRC Point Source Catalog Version 1 \\
X-ray & 2010 & 2PBC  & Second Palermo Swift-BAT hard X-ray catalog \\
Infrared & 2011 & WISE & Wide-field Infrared Survey Explorer catalog \\
X-ray & 2012 & CXO & Chandra Source Catalog Release 1.1 \\
Optical & 2012, 2014 & LQAC & Large Quasar Astrometric Catalog \\
Infrared & 2013 & SSTSL2 & Spitzer Space Telescope Source List - version 4.2 \\
X-ray & 2013 & 2MAXI & Monitor of All-sky X-ray Image 37-month catalog \\
Millimeter & 2013 & PLCKERC0(nn) & Planck Early Release Compact Source Catalog at (nn)GHz \\	 
Radio & 2014 & WB (WIBRaLS) & WISE Blazar-like Radio-loud Sources \\
Ultraviolet & 2016 & UVQS &	UV-bright Quasar Survey \\
Radio & 2016 & NVGRC & 	NVSS Giant Radio Sources Catalog \\
Gamma-ray & 2016 & MGRO & Milagro Gamma-Ray Observatory source list \\  
Optical & 2016- ... & Gaia DR(n) & Gaia Data Release (n) source \\
Gamma-ray & 2017 & 2HWC & HAWC Observatory first catalog \\
\enddata
\vspace{-7mm}
%
\end{deluxetable}

\end{appendix}

\setcounter{table}{0}
\begin{deluxetable}{lrr}
\setlength{\tabcolsep}{0.04in}
\tablewidth{0pt}
\tabletypesize{\scriptsize}
\tablecaption{Catalogs Used for the Bayesian Association Method \label{tab:catalogs}}
\tablehead{
\colhead{Name} & 
\colhead{Objects\tablenotemark{a}} & 
\colhead{Ref.}
}
\startdata
BZCAT (Blazars) & 3561  & \citet{bzcat} \\
BL Lac & 1371  & \citet{AGNcatalog} \\
AGN & 10066  & \citet{AGNcatalog} \\
QSO & 129,853  & \citet{AGNcatalog} \\
Seyfert galaxies & 27651  & \citet{AGNcatalog} \\
Narrow-line Seyfert galaxies & 18  &  \citet{NLSy1catalog_Berton} \\
Narrow-line Seyfert galaxies & 556  & \citet{NLSy1catalog} \\
FRICAT (Radiogalaxies)& 233 & \citet{FRICAT}\\ 
FRIICAT (Radiogalaxies) & 123 & \citet{FRIICAT}\\ 
Giant Radio Source & 349 & \citet{GRSKcatalog} \\
2WHSP & $~$1691 & \citet{2WHSP} \\
{\sl WISE} blazar catalog & 12319 & \citet{WISE}\\
Radio Fundamental Catalog (2019a) & {\bf 15740} & \url{http://astrogeo.org/rfc} \\
CGRaBS & 1625 & \citet{CGRaBS} \\
CRATES & 11499  & \citet{CRATES} \\
ATCA 20 GHz southern sky survey & 5890  & \citet{AT20G} \\
\enddata
\tablenotetext{a}{Number of objects in the catalog.}
\end{deluxetable}

\begin{table}
\caption{Comparison of association methods  in terms of total number of associations, $N_{\mathrm{assoc}}$, estimated number of false associations, $N_{\mathrm{false}}$,  and number of sources  associated only via a given method, $N{\mathrm{_S}}$.\label{tab:assoc}} 

\begin{center}
\begin{tabular}{|l|cc|ccc|ccc|}
   \hline
   Sample  & \multicolumn{2}{|c|}{All Methods }               & \multicolumn{3}{|c|}{ Bayesian Method} & \multicolumn{3}{|c|}{ LR Method} \\
   &  $N_{\mathrm{assoc}}$ & \multicolumn{1}{c|}{$N_{\mathrm{false}}$} &    \multicolumn{1}{|c}{$N_{\mathrm{assoc}}$} & \multicolumn{1}{c}{$N_{\mathrm{false}}$} &  \multicolumn{1}{c|}{$N_{\mathrm{S}}$ }  &   \multicolumn{1}{|c}{$N_{\mathrm{assoc}}$} & \multicolumn{1}{c}{$N_{\mathrm{false}}$} &  \multicolumn{1}{c|}{$N_{\mathrm{S}}$}   \\
   \hline
 Full Sample & \nsrc & \probtot & \nbay & \probbay & \nbayonly & \nlr & \problr & \nlronly \\
 Clean Sample& \nsrcc & \probtotc &\nbayc  &  \probbayc & \nbayonlyc & \nlrc & \problrc & \nlronlyc \\
\hline
\end{tabular}
\end{center}
\end{table}
\begin{deluxetable}{lccl}
\setlength{\tabcolsep}{0.04in}
\tablewidth{0pt}
\tabletypesize{\scriptsize}
\tablecaption{4LAC FITS Format \label{tab:description}}
\tablehead{
\colhead{Column} &
\colhead{Format} &
\colhead{Unit} &
\colhead{Description}
}
\startdata
Source\_Name & 18A & \nodata & Source name 4FGL JHHMM.m+DDMMa\tablenotemark{a} \\
RAJ2000 & E & deg & Right Ascension \\
DEJ2000 & E & deg & Declination \\
GLON & E & deg & Galactic Longitude \\
GLAT & E & deg & Galactic Latitude \\
Signif\_Avg & E & \nodata & Source significance in $\sigma$ units over the 50~MeV to 1~TeV band \\
Flux1000 & E & cm$^{-2}$ s$^{-1}$ & Integral photon flux from 1 to 100~GeV \\
Unc\_Flux1000 & E & cm$^{-2}$ s$^{-1}$ & $1\sigma$ error on integral photon flux from 1 to 100~GeV \\
Energy\_Flux100 & E & erg cm$^{-2}$ s$^{-1}$ & Energy flux from 100~MeV to 100~GeV obtained by spectral fitting \\
Unc\_Energy\_Flux100 & E & erg cm$^{-2}$ s$^{-1}$ & $1\sigma$  error on energy flux from 100~MeV to 100~GeV \\
SpectrumType & 17A & \nodata & Spectral type in the global model (PowerLaw, LogParabola, \\
& & &  PLSuperExpCutoff) \\
PL\_Index & E & \nodata & Photon index when fitting with PowerLaw \\
Unc\_PL\_Index & E & \nodata & $1\sigma$ error on PL\_Index \\
Pivot\_Energy & E & MeV & Pivot Energy \\
LP\_Index & E & \nodata & Photon index at Pivot\_Energy ($\alpha$) when fitting with LogParabola \\
Unc\_LP\_Index & E & \nodata & $1\sigma$ error on LP\_Index \\
LP\_beta & E & \nodata & Curvature parameter ($\beta$) when fitting with LogParabola \\
Unc\_LP\_beta & E & \nodata & $1\sigma$ error on LP\_beta \\
Flags & I & \nodata & Analysis flags \\
CLASS & 6A & \nodata & Class designation for associated source \\
ASSOC1 & 30A & \nodata & Name of identified or likely associated source \\
ASSOC\_PROB\_BAY & E & \nodata & Probability of association according to \\
& & & the Bayesian method \\
ASSOC\_PROB\_LR & E & \nodata & Probability of association according to \\
& & & the Likelihood Ratio method \\
Counterpart\_Catalog & 10A & & Counterpart catalog driving the association \\ 
RA\_Counterpart & D & deg & Right Ascension of the counterpart ASSOC1 \\
DEC\_Counterpart & D & deg & Declination of the counterpart ASSOC1 \\
Unc\_Counterpart & E & deg & 95\% precision of the counterpart localization \\
VLBI\_Counterpart   & 14A &  \nodata & Name of the VLBI counterpart \\ 
Gaia\_Counterpart   & 29A &  \nodata & Name of the Gaia counterpart established via the VLBI position \\
Gaia\_G\_Magnitude & E & \nodata & Gaia Magnitude at the G band\\
Redshift   &  E & \nodata  & Redshift \\
SED\_class  &  6A & \nodata & SED-based class\\
nu\_syn &  E & Hz & synchrotron-peak frequency  in observer frame \\
nuFnu\_syn  & E  & erg cm$^{-2}$ s$^{-1}$ & Spectral energy distribution at synchrotron-peak frequency \\
Variability\_Index & E &  \nodata & Variability index \\
Frac\_Variability  & E &  \nodata & Fractional variability  \\
Unc\_Frac\_Variability   & E &  \nodata & 1$\sigma$ error on fractional  variability \\
Highest\_energy &  E & GeV &  energy of the highest-energy ULTRACLEANVETO photon \\
& & & with association probability $P>0.95$ \\
\enddata
\end{deluxetable}

\begin{deluxetable}{lrrr}
\tablecolumns{4}
\tabletypesize{\footnotesize}
\tablecaption{\label{tab:census}Census of sources}
\tablewidth{0pt}
\tablehead{
\colhead{AGN type}&
\colhead{Entire 4LAC}&
\colhead{4LAC Clean Sample\tablenotemark{a}}&
\colhead{Low-latitude sample}
}
\startdata
{\bf All}&{\bf \nsrc}&{\bf \nsrcc }&{\bf \nlowlat}\\
\\
{\bf FSRQ}&{\bf \nhighlatf}&{\bf \nhighlatfc}&{\bf \nlowlatf}\\
{\ldots}LSP& \nhighlataa & \nhighlatcaa & \nlowlataa\\
{\ldots}ISP& \nhighlatab &\nhighlatcab & \nlowlatab\\
{\ldots}HSP& \nhighlatac &\nhighlatcac &\nlowlatac\\
{\ldots}no SED classification & \nhighlatad &\nhighlatcad & \nlowlatad\\
\\
{\bf BL Lac}&{\bf \nhighlatbl}&{\bf \nhighlatblc}&{\bf \nlowlatbl}\\
{\ldots}LSP& \nhighlatba & \nhighlatcba &\nlowlatba\\
{\ldots}ISP& \nhighlatbb &\nhighlatcbb &\nlowlatbb\\
{\ldots}HSP& \nhighlatbc &\nhighlatcbc & \nlowlatbc\\
{\ldots}no SED classification & \nhighlatbd & \nhighlatcbd & \nlowlatbd\\
\\
{\bf Blazar of Unknown type}&{\bf \nhighlatu}&{\bf \nhighlatuc}&{\bf \nlowlatu}\\
{\ldots}LSP&  \nhighlatca& \nhighlatcca & \nlowlatca\\
{\ldots}ISP& \nhighlatcb &\nhighlatccb & \nlowlatcb\\
{\ldots}HSP& \nhighlatcc &\nhighlatccc & \nlowlatcc\\
{\ldots}no SED classification & \nhighlatcd &\nhighlatccd & \nlowlatcd\\
\\
{\bf Non-blazar AGN}&{\bf \nhighlatag}&{\bf \nhighlatagc}&{\bf \nlowlatag}\\
{\ldots}NLSy1 & \nnlsy & \nnlsyc  & \nnlsyl \\
{\ldots}RG & \nrdg & \nrdgc & \nrdgl\\
{\ldots}CSS & \ncss & \ncssc & \ncssl\\
{\ldots}SSRQ & \nssrq & \nssrqc & \nssrql\\
{\ldots}SEY & 0 & 0 & 1\\
{\ldots}Other AGN & \nagn& \nagnc& \nagnl\\
\\
\enddata
\tablenotetext{a}{Sources with single counterparts and without analysis flags. See Section~\ref{sec:assocresults} for the definitions of this sample.}
\end{deluxetable}

\begin{deluxetable}{lllc}
\rotate
\tablewidth{0pt}
\tabletypesize{\footnotesize}
\tablecolumns{2}
\tablecaption{3LAC sources not present in 4LAC}
\tablehead{
\colhead{3FGL Name} &
\colhead{Counterpart} &
\colhead{Class} &
\colhead{4FGL$\dagger$} \\
}
\startdata
 J0127.9+2551 & 4C +25.05                  & fsrq    & M \\
 J0135.0+6927 & TXS 0130+691               & bcu & M \\
 J0211.7+5402 & TXS 0207+538               & bcu & M \\
 J0216.1-7016 & PMN J0215-7014             & bcu & M \\
 J0217.3+6209 & TXS 0213+619               & bcu & M \\
 J0223.5+6313 & TXS 0219+628               & bcu & M \\
 J0228.5+6703 & GB6 J0229+6706             & bcu & M \\
 J0302.0+5335 & GB6 J0302+5331             & bcu & M \\
 J0336.9-1304 & PKS 0334-131               & fsrq    & M \\
 J0351.1+0128 & TXS 0348+013               & ssrq    & M \\
 J0512.2+2918 & B2 0509+29                 & bcu & M \\
 J0514.4+5603 & TXS 0510+559               & fsrq    & M \\
 J0517.4+4540 & 4C +45.08                  & fsrq    & M \\
 J0528.3+1815 & 1RXS J052829.6+181657      & bcu & M \\
 J0618.9-1138 & TXS 0616-116               & bcu & M \\
 J0627.9-1517 & NVSS J062753-152003        & bcu & M \\
 J0730.3+6720 & GB6 J0731+6718             & fsrq    & M \\
 J0742.4-8133 & SUMSS J074220-813139       & bcu & M \\
 J0744.1-3804 & PMN J0743-3804             & bcu & M \\
 J0904.9+2739 & GB6 J0905+2748A            & fsrq    & M \\
 J0928.7+7300 & GB6 J0929+7304             & bcu & M \\
 J0956.7-6441 & AT20G J095612-643928       & bcu & M \\
 J1005.0-4959 & PMN J1006-5018             & bcu & M \\
 J1016.0-0635 & NVSS J101626-063624        & bcu & M \\
 J1024.8+0105 & PMN J1024+0056             & bcu & M \\
 J1037.4-3742 & PKS 1034-374               & fsrq    & M \\
 J1123.2-6415 & AT20G J112319-641735       & bcu & M \\
 J1205.4+0412 & MG1 J120448+0408           & fsrq    & M \\
 J1218.5+6912 & NVSS J122044+690522        & bcu & M \\
 J1326.1+2931 & TXS 1323+298               & bll     & M \\
 J1330.0-3818 & Tol 1326-379               & fsrq    & M \\
 J1356.3-4029 & SUMSS J135625-402820       & bcu & M \\
 J1415.0-1001 & PKS B1412-096              & fsrq    & M \\
 J1509.9-2951 & TXS 1507-296               & bcu & M \\
 J1513.1-1014 & PKS 1511-100               & fsrq    & M \\
 J1514.1+2940 & MG2 J151421+2930           & fsrq    & M \\
 J1536.6+8331 & NVSS J153556+832614        & bcu & M \\
 J1541.8+1105 & MG1 J154207+1110           & fsrq    & M \\
 J1621.1-2331 & PKS 1617-235               & agn     & M \\
 J1645.2-5747 & AT20G J164513-575122       & bcu & M \\
 J1648.5-4829 & PMN J1648-4826             & bcu & M \\
 J1723.5-5609 & PMN J1723-5614             & bcu & M \\
 J1747.1+0139 & PMN J1746+0141             & bcu & M \\
 J1757.4+6536 & 7C 1757+6536               & bcu & M \\
 J1804.1+0341 & TXS 1801+036               & fsrq    & M \\
 J1819.1+4259 & NVSS J181927+425800        & bcu & M \\
 J1822.1-7051 & PMN J1823-7056             & bcu & M \\
 J1949.4-6140 & PMN J1949-6137             & bcu & M \\
 J2107.7-4822 & PMN J2107-4827             & bcu & M \\
 J2151.6-2744 & PMN J2151-2742             & fsrq    & M \\
 J2203.7+3143 & 4C +31.63                  & fsrq    & M \\
 J2236.2-5049 & SUMSS J223605-505521       & bcu & M \\
 J2246.2+1547 & NVSS J224604+154437        & bcu & M \\
 J2305.3-4219 & SUMSS J230512-421859       & bcu & M \\
 J2343.6+1551 & MG1 J234342+1542           & fsrq    & M \\
 J0003.8-1151 & PKS 0001-121               & bcu & D \\
 J0009.1+0630 & GB6 J0009+0625             & bll     & D \\
 J0059.1-5701 & PKS 0056-572               & bcu & L \\
 J0203.6+3043 & B2 0200+30                 & fsrq    & D \\
 J0426.6+0459 & 4C +04.15                  & bcu & L \\
 J0442.6-0017 & 1RXS J044229.8-001823      & bll     & D \\
 J0447.8-2119 & PKS 0446-212               & fsrq    & L \\
 J0515.3-4557 & PMN J0514-4554             & bcu & D \\
 J0526.0+4253 & NVSS J052520+425520        & bcu & C \\
 J0542.2-8737 & SUMSS J054923-874001       & bcu & L \\
 J0618.0+7819 & 1REX J061757+7816.1        & fsrq    & C \\
 J0647.1-4415 & SUMSS J064648-441929       & bcu & C \\
 J0712.2-6436 & MRC 0712-643               & bcu & L \\
 J0744.8-4028 & PMN J0744-4032             & bcu & L \\
 J0807.9+4946 & SDSS J080754.50+494627.6   & fsrq    & D \\
 J0824.9+3916 & 4C +39.23B                 & css     & D \\
 J0825.4-0213 & PMN J0825-0204             & bcu & L \\
 J0934.1+3933 & 3C 221                     & rdg     & D \\
 J1007.4-3334 & TXS 1005-333               & bcu & D \\
 J1010.8-0158 & NVSS J101051-020227        & fsrq    & D \\
 J1018.1+1904 & MG1 J101810+1903           & bcu & D \\
 J1048.6+2338 & NVSS J104900+233821        & bll     & C \\
 J1101.5+4106 & B3 1058+413                & bcu & D \\
 J1146.8+3958 & NVSS J114653+395751        & bcu & D \\
 J1207.6-4537 & PMN J1207-4531             & bcu & L \\
 J1244.1+1615 & 3C 275.1                   & ssrq    & D \\
 J1256.7+5328 & TXS 1254+538               & bcu & L \\
 J1300.2+1416 & NVSS J130041+141728        & bcu & D \\
 J1322.8-0938 & PMN J1323-0943             & bcu & D \\
 J1451.2+6355 & TXS 1450+641               & bcu & D \\
 J1514.8+4446 & NVSS J151436+445003        & fsrq    & L \\
 J1554.4+2010 & 1ES 1552+203               & bll     & L \\
 J1617.3-2519 & TXS 1613-251               & agn     & D \\
 J1625.9+4125 & 4C +41.32                  & fsrq    & D \\
 J1908.8-0130 & NVSS J190836-012642        & bcu & L \\
 J2036.8-2830 & PMN J2036-2830             & fsrq    & L \\
 J2110.3+3540 & B2 2107+35A                & bcu & L \\
 J2348.4-5100 & SUMSS J234852-510311       & bcu & L \\

\hline
\enddata
\tablecomments{$\dagger$ M: missing gamma-ray source in 4FGL.  L: 3FGL source present in 4FGL but now unassociated.  C: 3FGL source present in 4FGL but now associated with a  non-AGN counterpart. D: duplicate counterpart in 3LAC now missing in 4LAC.
}
\label{tab:miss}
\end{deluxetable}

\begin{deluxetable}{lllll}
\setlength{\tabcolsep}{0.025in}
\tabletypesize{\scriptsize}
\tablewidth{350pt}
\tablecaption{Non-blazar objects and misaligned AGNs}
\tablehead{
\colhead{4FGL Name} &
\colhead{Name} &
\colhead{Type} &
\colhead{Photon Index} &
\colhead{Redshift / Distance (Mpc)}  }
\startdata
J0009.7-3217 & IC 1531 & rdg & 2.2$\pm$0.14& 93.4* \\ 
J0013.6+4051 & 4C +40.01 & agn & 2.21$\pm$0.14& 0.255 \\ 
J0038.7-0204 & 3C 17 & rdg & 2.81$\pm$0.11& 0.22 \\ 
J0057.7+3023 & NGC 315 & rdg & 2.35$\pm$0.11& 0.016 \\ 
J0237.7+0206 & PKS 0235+017 & rdg & 2.17$\pm$0.18& 0.022 \\ 
J0308.4+0407 & NGC 1218 & rdg & 2.0$\pm$0.06& 0.029 \\ 
J0312.9+4119 & B3 0309+411B & rdg & 2.47$\pm$0.19& 0.136 \\ 
J0316.8+4120 & IC 310 & rdg & 1.78$\pm$0.18& 0.019 \\ 
J0319.8+4130 & NGC 1275 & rdg & 2.12$\pm$0.01& 0.69* \\ 
J0322.6-3712e & Fornax A & rdg & 2.05$\pm$0.07& 17.8* \\ 
J0324.8+3412 & 1H 0323+342 & nlsy1 & 2.82$\pm$0.04& 0.061 \\ 
J0334.3+3920 & 4C +39.12 & rdg & 1.9$\pm$0.13& 0.021 \\ 
J0433.0+0522 & 3C 120 & rdg & 2.72$\pm$0.05& 0.033 \\ 
J0519.6-4544 & Pictor A & rdg & 2.46$\pm$0.13& 0.035 \\ 
J0521.2+1637 & 3C 138 & css & 2.37$\pm$0.13& 0.759 \\ 
J0522.9-3628 & PKS 0521-36 & agn & 2.45$\pm$0.01& 0.056 \\ 
J0627.0-3529 & PKS 0625-35 & rdg & 1.9$\pm$0.04& 0.055 \\ 
J0708.9+4839 & NGC 2329 & rdg & 1.95$\pm$0.18& 0.019 \\ 
J0758.7+3746 & NGC 2484 & rdg & 2.01$\pm$0.16& 171* \\ 
J0840.8+1317 & 3C 207 & ssrq & 2.48$\pm$0.1& 0.681 \\ 
J0850.0+5108 & SBS 0846+513 & nlsy1 & 2.27$\pm$0.02& 0.583 \\ 
J0858.1+1405 & 3C 212 & ssrq & 2.52$\pm$0.15& 1.048 \\ 
J0910.0+4257 & 3C 216 & css & 2.52$\pm$0.11& 0.67 \\ 
J0931.9+6737 & NGC 2892 & rdg & 2.23$\pm$0.06& 0.023 \\ 
J0948.9+0022 & PMN J0948+0022 & nlsy1 & 2.64$\pm$0.02& 0.585 \\ 
J1012.7+4228 & B3 1009+427 & agn & 1.76$\pm$0.09& 0.365 \\ 
J1116.6+2915 & B2 1113+29 & rdg & 1.44$\pm$0.24& 0.047 \\ 
J1118.2-0415 & PMN J1118-0413 & agn & 2.64$\pm$0.08& \nodata \\ 
J1144.9+1937 & 3C 264 & rdg & 1.94$\pm$0.1& 0.022 \\ 
J1149.0+5924 & NGC 3894 & rdg & 2.06$\pm$0.12& 46.9* \\ 
J1230.8+1223 & M 87 & rdg & 2.06$\pm$0.04& 16.5* \\ 
J1305.3+5118 & IERS B1303+515 & nlsy1 & 2.85$\pm$0.17& 0.788 \\ 
J1306.3+1113 & TXS 1303+114 & rdg & 1.95$\pm$0.18& 0.086 \\ 
J1306.7-2148 & PKS 1304-215 & rdg & 2.13$\pm$0.09& 0.126 \\ 
J1325.5-4300 & Cen A & rdg & 2.65$\pm$0.02& 3.8* \\ 
J1331.0+3032 & 3C 286 & css & 2.41$\pm$0.14& 0.85 \\ 
J1356.2-1726 & PKS B1353-171 & agn & 2.08$\pm$0.15& 0.075 \\ 
J1443.1+5201 & 3C 303 & rdg & 1.98$\pm$0.15& 0.141 \\ 
J1443.1+4728 & B3 1441+476 & nlsy1 & 2.56$\pm$0.11& 0.705 \\ 
J1449.5+2746 & B2 1447+27 & rdg & 1.54$\pm$0.18& 0.031 \\ 
J1449.7-0910 & 1RXS J144942.2-091018 & agn & 2.04$\pm$0.18& \nodata \\ 
J1459.0+7140 & 3C 309.1 & css & 2.45$\pm$0.09& 0.91 \\ 
J1505.0+0326 & PKS 1502+036 & nlsy1 & 2.59$\pm$0.04& 0.409 \\ 
J1516.5+0015 & PKS 1514+00 & rdg & 2.59$\pm$0.11& 0.052 \\ 
J1518.6+0614 & TXS 1516+064 & rdg & 1.86$\pm$0.17& 0.102 \\ 
J1521.1+0421 & PKS B1518+045 & rdg & 2.06$\pm$0.15& 0.052 \\ 
J1543.6+0452 & CGCG 050-083 & agn & 1.87$\pm$0.08& 0.04 \\ 
J1630.6+8234 & NGC 6251 & rdg & 2.35$\pm$0.03& 98.2* \\ 
J1644.9+2620 & MG2 J164443+2618 & nlsy1 & 2.78$\pm$0.1& 0.144 \\ 
J1724.2-6501 & NGC 6328 & rdg & 2.49$\pm$0.18& 0.014 \\ 
J1829.5+4845 & 3C 380 & css & 2.43$\pm$0.03& 0.695 \\ 
J1843.4-4835 & PKS 1839-48 & rdg & 1.99$\pm$0.17& 0.111 \\ 
J2007.9-4432 & PKS 2004-447 & nlsy1 & 2.6$\pm$0.05& 0.24 \\ 
J2114.8+2026 & TXS 2112+202 & agn & 2.13$\pm$0.16& \nodata \\ 
J2118.8-0723 & TXS 2116-077 & nlsy1 & 2.83$\pm$0.15& 0.26 \\ 
J2156.0-6942 & PKS 2153-69 & rdg & 2.83$\pm$0.11& 0.028 \\ 
J2227.9-3031 & PKS 2225-308 & rdg & 1.99$\pm$0.17& 0.056 \\ 
J2302.8-1841 & PKS 2300-18 & rdg & 2.17$\pm$0.15& 0.129 \\ 
J2326.9-0201 & PKS 2324-02 & rdg & 2.44$\pm$0.14& 0.188 \\ 
J2329.7-2118 & PKS 2327-215 & rdg & 2.45$\pm$0.16& 0.031 \\ 
J2334.9-2346 & PKS 2331-240 & agn & 2.42$\pm$0.12& 0.048 \\ 
J2338.1+0325 & PKS 2335+03 & agn & 2.36$\pm$0.15& 0.27 \\ 
J2341.8-2917 & PKS 2338-295 & rdg & 2.24$\pm$0.15& 0.052 \\ 

\hline
J0153.4+7114 & TXS 0149+710 & rdg & 1.9$\pm$0.11& 0.022 \\ 
J0418.2+3807 & 3C 111 & rdg & 2.71$\pm$0.06& 0.05 \\ 
J1236.9-7232 & PKS 1234-723 & rdg & 2.36$\pm$0.14& 0.024 \\ 
J1346.3-6026 & Cen B & rdg & 2.4$\pm$0.05& 0.013 \\ 
J1413.1-6519 & Circinus galaxy & sey & 2.25$\pm$0.1& 4.0* \\ 
J1824.7-3243 & PKS 1821-327 & agn & 2.23$\pm$0.12& 0.355 \\ 

\enddata
\tablecomments{
The table includes the  non-blazar objects and MAGNs at high latitudes (top) and low latitudes (bottom) associated with 4FGL sources (Cen A Core and Cen A Lobes constitute a single object). $^{*}$ indicates that the value is the  distance in Mpc.}
\label{tab:magn}
\end{deluxetable}

\clearpage

\begin{deluxetable}{lcccccc}
\setlength{\tabcolsep}{0.025in}
\tabletypesize{\scriptsize}
\tablewidth{0pt}
\tablecaption{Properties of the 4LAC VHE AGNs }
\tablehead{
\colhead{4FGL } &
\colhead{Name } &
\colhead{Source} &
\colhead{SED} &
\colhead{Redshift} &
\colhead{Photon} &
\colhead{Variability} 
\\
\colhead{Name}  &
\colhead{} &
\colhead{Class} &
\colhead{Type} &
\colhead{} &
\colhead{Index} &
\colhead{Index$^a$} 
\\
}
\startdata
J0013.9-1854 & RBS 0030 & bll & HSP &0.09 & 1.97$\pm$0.1 & 2.61\\
J0033.5-1921 & KUV 00311-1938 & bll & HSP &0.61 & 1.77$\pm$0.02 & 27.17\\
J0035.9+5950 & 1ES 0033+595$^*$ & bll & HSP &\nodata & 1.75$\pm$0.02 & 149.76\\
J0112.1+2245 & S2 0109+22 & bll & LSP &0.26 & 2.07$\pm$0.01 & 313.15\\
J0136.5+3906 & B3 0133+388 & bll & HSP &\nodata & 1.71$\pm$0.02 & 51.08\\
J0152.6+0147 & PMN J0152+0146 & bll & HSP &0.08 & 1.96$\pm$0.06 & 15.23\\
J0214.3+5145 & TXS 0210+515$^*$ & bll & HSP &0.05 & 1.88$\pm$0.09 & 11.2\\
J0221.1+3556 & B2 0218+357 & fsrq & LSP &0.94 & 2.29$\pm$0.01 & 3327.03\\
J0222.6+4302 & 3C 66A & bll & ISP &0.44 & 1.96$\pm$0.01 & 976.11\\
J0232.8+2018 & 1ES 0229+200 & bll & HSP &0.14 & 1.78$\pm$0.11 & 4.36\\
J0238.4-3116 & 1RXS J023832.6-311658 & bll & HSP &0.23 & 1.8$\pm$0.04 & 33.44\\
J0303.4-2407 & PKS 0301-243 & bll & HSP &0.27 & 1.9$\pm$0.02 & 253.95\\
J0316.8+4120 & IC 310 & rdg & ISP &0.02 & 1.78$\pm$0.18 & 19.86\\
J0319.8+4130 & NGC 1275 & rdg & LSP &0.02 & 2.12$\pm$0.01 & 1970.98\\
J0319.8+1845 & 1E 0317.0+1835 & bll & HSP &0.19 & 1.67$\pm$0.07 & 32.33\\
J0349.4-1159 & 1ES 0347-121 & bll & HSP &0.19 & 1.76$\pm$0.1 & 10.0\\
J0416.9+0105 & 1ES 0414+009 & bll & HSP &0.29 & 1.89$\pm$0.07 & 9.76\\
J0449.4-4350 & PKS 0447-439 & bll & HSP &0.2 & 1.85$\pm$0.01 & 55.2\\
J0507.9+6737 & 1ES 0502+675 & bll & HSP &0.42 & 1.58$\pm$0.03 & 39.18\\
J0509.4+0542 & TXS 0506+056 & bll & ISP &0.34 & 2.08$\pm$0.02 & 245.91\\
J0521.7+2112 & TXS 0518+211$^*$ & bll & HSP &0.11 & 1.92$\pm$0.01 & 682.16\\
J0550.5-3216 & PKS 0548-322 & bll & HSP &0.07 & 1.89$\pm$0.1 & 5.98\\
J0627.0-3529 & PKS 0625-35 & rdg & HSP &0.05 & 1.9$\pm$0.04 & 11.56\\
J0648.7+1516 & RX J0648.7+1516$^*$ & bll & HSP &0.18 & 1.7$\pm$0.04 & 3.2\\
J0650.7+2503 & 1ES 0647+250 & bll & HSP &0.2 & 1.74$\pm$0.02 & 117.43\\
J0710.4+5908 & 1H 0658+595 & bll & HSP &0.13 & 1.68$\pm$0.05 & 26.55\\
J0721.9+7120 & S5 0716+71 & bll & ISP &0.13 & 2.06$\pm$0.01 & 1554.68\\
J0733.4+5152 & NVSS J073326+515355 & bcu & HSP &0.06 & 1.8$\pm$0.1 & 14.97\\
J0739.2+0137 & PKS 0736+01 & fsrq & LSP &0.19 & 2.41$\pm$0.02 & 1983.59\\
J0809.8+5218 & 1ES 0806+524 & bll & HSP &0.14 & 1.88$\pm$0.02 & 290.4\\
J0847.2+1134 & RX J0847.1+1133 & bll & ISP &0.2 & 1.72$\pm$0.08 & 8.47\\
J0854.8+2006 & OJ 287 & bll & LSP &0.31 & 2.23$\pm$0.01 & 611.67\\
J0958.7+6534 & S4 0954+65 & bll & LSP &0.37 & 2.21$\pm$0.02 & 2012.73\\
J1010.2-3119 & 1RXS J101015.9-311909 & bll & HSP &0.14 & 1.75$\pm$0.07 & 34.16\\
J1015.0+4926 & 1H 1013+498 & bll & HSP &0.21 & 1.84$\pm$0.01 & 195.6\\
J1103.6-2329 & 1ES 1101-232 & bll & HSP &0.19 & 1.73$\pm$0.08 & 10.7\\
J1104.4+3812 & Mkn 421 & bll & HSP &0.03 & 1.78$\pm$0.01 & 1028.05\\
J1136.4+6736 & RX J1136.5+6737 & bll & HSP &0.14 & 1.75$\pm$0.05 & 23.5\\
J1136.4+7009 & Mkn 180 & bll & HSP &0.05 & 1.8$\pm$0.03 & 20.39\\
J1144.9+1937 & 3C 264 & rdg & HSP &0.02 & 1.94$\pm$0.1 & 4.15\\
J1159.5+2914 & Ton 599 & fsrq & LSP &0.73 & 2.26$\pm$0.01 & 1391.63\\
J1217.9+3007 & B2 1215+30 & bll & HSP &0.13 & 1.95$\pm$0.01 & 396.98\\
J1221.3+3010 & PG 1218+304 & bll & HSP &0.18 & 1.71$\pm$0.02 & 44.28\\
J1221.5+2814 & W Comae & bll & ISP &0.1 & 2.16$\pm$0.02 & 243.32\\
J1224.4+2436 & MS 1221.8+2452 & bll & HSP &0.22 & 1.89$\pm$0.04 & 148.48\\
J1224.9+2122 & 4C +21.35 & fsrq & LSP &0.43 & 2.33$\pm$0.01 & 17566.6\\
J1230.2+2517 & ON 246 & bll & ISP &0.14 & 2.09$\pm$0.02 & 1651.75\\
J1230.8+1223 & M 87 & rdg & ISP &0.01 & 2.06$\pm$0.04 & 16.98\\
J1256.1-0547 & 3C 279 & fsrq & LSP &0.54 & 2.34$\pm$0.01 & 5667.24\\
J1315.0-4236 & MS 13121-4221 & bll & HSP &0.1 & 1.72$\pm$0.1 & 6.61\\
J1325.5-4300 & Cen A & rdg & LSP &0.01 & 2.65$\pm$0.02 & 8.25\\
J1427.0+2348 & PKS 1424+240 & bll & HSP &0.6 & 1.82$\pm$0.01 & 205.42\\
J1428.5+4240 & H 1426+428 & bll & HSP &0.13 & 1.66$\pm$0.05 & 14.91\\
J1442.7+1200 & 1ES 1440+122 & bll & HSP &0.16 & 1.8$\pm$0.07 & 10.06\\
J1443.9-3908 & PKS 1440-389 & bll & HSP &0.07 & 1.82$\pm$0.02 & 22.05\\
J1443.9+2501 & PKS 1441+25 & fsrq & LSP &0.94 & 2.08$\pm$0.02 & 2858.38\\
J1512.8-0906 & PKS 1510-089 & fsrq & LSP &0.36 & 2.38$\pm$0.01 & 4421.04\\
J1517.7-2422 & AP Librae & bll & LSP &0.05 & 2.12$\pm$0.02 & 90.89\\
J1518.0-2731 & TXS 1515-273 & bll & HSP &\nodata & 2.06$\pm$0.05 & 57.84\\
J1555.7+1111 & PG 1553+113 & bll & HSP &0.36 & 1.68$\pm$0.01 & 74.97\\
J1653.8+3945 & Mkn 501 & bll & HSP &0.03 & 1.75$\pm$0.01 & 292.85\\
J1725.0+1152 & 1H 1720+117 & bll & HSP &0.18 & 1.86$\pm$0.02 & 13.33\\
J1728.3+5013 & I Zw 187 & bll & HSP &0.05 & 1.78$\pm$0.03 & 164.28\\
J1744.0+1935 & S3 1741+19 & bll & HSP &0.08 & 1.93$\pm$0.05 & 10.72\\
J1751.5+0938 & OT 081 & bll & LSP &0.32 & 2.26$\pm$0.02 & 884.58\\
J1944.0+2117 & MG2 J194359+2118$^*$ & bcu &  &\nodata & 1.53$\pm$0.09 & 15.42\\
J2000.0+6508 & 1ES 1959+650 & bll & HSP &0.05 & 1.82$\pm$0.01 & 1052.41\\
J2001.2+4353 & MG4 J200112+4352$^*$ & bll & HSP &\nodata & 1.95$\pm$0.02 & 1027.21\\
J2009.4-4849 & PKS 2005-489 & bll & HSP &0.07 & 1.83$\pm$0.02 & 135.08\\
J2039.5+5218 & 1ES 2037+521$^*$ & bll & HSP &0.05 & 1.88$\pm$0.09 & 4.58\\
J2056.7+4939 & RGB J2056+496$^*$ & bcu & HSP &\nodata & 1.85$\pm$0.04 & 23.86\\
J2158.8-3013 & PKS 2155-304 & bll & HSP &0.12 & 1.85$\pm$0.01 & 646.95\\
J2202.7+4216 & BL Lac & bll & LSP &0.07 & 2.23$\pm$0.01 & 2474.03\\
J2243.9+2021 & RGB J2243+203 & bll & HSP &\nodata & 1.86$\pm$0.02 & 116.39\\
J2250.0+3825 & B3 2247+381 & bll & HSP &0.12 & 1.72$\pm$0.06 & 24.53\\
J2324.7-4041 & 1ES 2322-409 & bll & HSP &0.17 & 1.78$\pm$0.05 & 49.44\\
J2347.0+5141 & 1ES 2344+514$^*$ & bll & HSP &0.04 & 1.81$\pm$0.02 & 56.97\\
J2359.0-3038 & H 2356-309 & bll & HSP &0.17 & 1.93$\pm$0.07 & 1.77\\

\enddata
\tablecomments{* refers to low-latitude sources (not in 4LAC). \\ $^a$ A variability index greater than 18.47 indicates that the source is variable at a significance greater than 99\%. }
\label{tab:GeVTeV}
\end{deluxetable}


\end{document}